%% file: main.tex
\documentclass[conference,a4paper]{IEEEtran}
\IEEEoverridecommandlockouts
\usepackage[utf8]{inputenc} 
\usepackage[T1]{fontenc}
\usepackage{url}              
\usepackage{cite}             

\usepackage[cmex10]{amsmath}  
\usepackage{mleftright}       
\mleftright                   

\usepackage{graphicx}         
\usepackage{booktabs}         

\usepackage[caption=false,font=footnotesize]{subfig}

\usepackage{stfloats}

\usepackage[utf8]{inputenc}
\usepackage{relsize}
\usepackage{amsfonts, amssymb, cuted}
\usepackage{cleveref}
\usepackage{mathtools}
\usepackage{algorithm}
\usepackage[noend]{algpseudocode}
\usepackage{xcolor}
\usepackage{soul}
\usepackage{nicefrac}
\usepackage{enumitem}
\usepackage{acro}
\usepackage{pifont}

\title{Cache-Aided Communications in MISO Networks with Dynamic User Behavior: A Universal Solution}
\author{\IEEEauthorblockN{Milad Abolpour, MohammadJavad Salehi, and Antti T\"olli} \\
\IEEEauthorblockA{
    Centre for Wireless Communications, University of Oulu, 90570 Oulu, Finland \\
    \textrm{E-mail: \{firstname.lastname\}@oulu.fi}
}
\thanks{This research has been supported by the Academy of Finland, 6G Flagship program under Grant 346208, 343586 (CAMAIDE), and by the Finnish-American Research and Innovation Accelerator (FARIA).}
}

\begin{document}
\input{commands.tex}
\input{MyCommands.tex}
\maketitle

\begin{abstract}
A practical barrier to the implementation of cache-aided networks is dynamic and unpredictable user behavior. 
In dynamic setups, users can freely depart and enter the network at any moment.  The shared caching concept has the potential to handle this issue by assigning $K$ users to $P$ caching profiles, where all $\eta_{p}$ users  assigned to profile $p$ store the same cache content defined by that profile. The existing schemes, however, cannot be applied in general
and are not dynamic in the true sense 
as they put constraints on the transmitter-side spatial multiplexing gain $\alpha$. Specifically, they work only if $\alpha \leq \min_{p} \eta_{p}$ or $\alpha \geq \hat{\eta}$, where in the latter case, 
$\gamma$ is the normalized cache size of each user, $\hat{\eta}$ is an arbitrary parameter satisfying $1 \leq \hat{\eta} \leq \max_{p} \eta_{p}$, and the extra condition of $\alpha \geq K\gamma$ should also be met.
%
%
In this work, we propose a universal caching scheme based on the same shared-cache model that can be applied to any dynamic setup,
extending the working region of existing schemes to networks with $\min_{p} \eta_{p} \leq \alpha \leq \hat{\eta}$ and removing any other constraints of existing schemes.
%
%
%
We also
derive the closed-form expressions for the achievable degrees-of-freedom (DoF)
of the proposed scheme and show that it achieves the optimal DoF for uniform user distributions.
Notably, it is the first scheme to achieve the optimal DoF of
$K\gamma+\alpha$ for networks with uniform user distribution, $\alpha > \hat{\eta}$, and non-integer $\frac{\alpha}{\hat{\eta}}$, without imposing any other constraints.
Finally, we use numerical simulations to 
assess how non-uniform user distribution impacts the DoF performance and illustrate that the proposed scheme provides a noticeable improvement over unicasting for uneven distributions.
\end{abstract}
\begin{IEEEkeywords}
coded caching; shared caching; dynamic networks; multi-antenna communications
\end{IEEEkeywords}

\section{Introduction}
The increasing volume and diversity of multimedia content require wireless networks to be enhanced to serve users at higher data rates and with lower latency~\cite{cisco2018cisco}, while network providers must further develop infrastructure in anticipation of evolving applications such as wireless immersive viewing~\cite{mahmoodi2021non,salehi2022enhancing}. To facilitate the efficient delivery of such multimedia content, coded caching (CC) has been proposed to increase the data rates by leveraging the cache memory across the network as a communication resource~\cite{maddah2014fundamental}. Accordingly, incorporating CC into a single-stream downlink network boosts the achievable rate by a multiplicative factor proportional to the cumulative cache size in the entire network via multicasting carefully designed codewords to different user groups. In light of the significance of multi-antenna connectivity in deploying next-generation networks~\cite{cisco2018cisco}, various works have studied the performance of cache-aided multi-input single-output (MISO) configurations~\cite{shariatpanahi2017multi,zhao2020multi,shariatpanahi2016multi,tolli2017multi,shariatpanahi2018physical,lampiris2019bridgingb}. For instance,~\cite{tolli2017multi} and~\cite{salehi2019subpacketization} discussed the design of optimized beamformers in finite signal-to-noise-ratio (SNR), and~\cite{mahmoodi2021non} explored the capability of CC to cope with location-dependent file request applications. 

In practice, however, practically achievable CC gains are constrained by the subpacketization process \cite{salehi2020lowcomplexity,yan2017placement,chittoor2020subexponential}. That is, in a network with $K$ users, each file should be split into many smaller parts, the number of which grows exponentially with $K$. A promising way to overcome this impediment is to use the shared caching concept, where there exist $P \le K$ \emph{caching profiles}, and $\eta_{p}$ users are assigned to profile  $p \in \left\lbrace 1,\cdots,P \right\rbrace$. Even though with this concept, multiple users  with a cache ratio of $\gamma$ could be assigned to the same profile and cache exactly the same data, in~\cite{parrinello2020extending}, it is shown that in MISO setups with $\alpha \geq \frac{K}{P}$, the scaling factor in the degrees-of-freedom (DoF) could be the same as the case of dedicated users' caches, i.e., $K\gamma+\alpha$, where $\alpha$ is the spatial multiplexing gain. However, for a shared-cache MISO setup with $\alpha \leq  \frac{K}{P}$, the optimal DoF is $\alpha \left( 1+P\gamma \right)$ \cite{parrinello2019fundamental}.

Interestingly, shared caching can also address another critical issue with coded caching schemes: handling networks with a dynamic population of users departing and entering the network at any time. The problem with conventional CC schemes is that they require the placement phase to be designed based on the number of users known a priori. By contrast, this problem is alleviated with the shared-cache model since the cache placement phase is built upon knowledge of the number of profiles $P$, and not the number of users $K$. Accordingly, in some cache-aided  scenarios, such as extended reality applications \cite{salehi2022enhancing}, the server is aware of the cache ratio of users rather than the number of existing users. In this sense, the authors in  \cite{Abolpour2022CodedNetworks} and \cite{salehi2021low}, apply the shared caching idea to address the dynamicity issue. In this method, the server only needs to know the cache ratio of users to determine the number of caching profiles and design the content placement phase. Although shared caching is crucial for managing dynamic conditions, the existing models are not dynamic in the true sense and only support two regions: \textit{1)} $\alpha \leq \min_{p} \eta_{p}$~\cite{parrinello2019fundamental}, and  \textit{2)} $\alpha \geq \hat{\eta}$ with $\alpha \geq K\gamma$ and arbitrary $\hat{\eta}$ satisfying $1 \leq \hat{\eta} \leq \max_{p} \eta_{p}$ \cite{Abolpour2022CodedNetworks,salehi2021low}.
Therefore, a universal shared-cache setup that supports any user-to-profile association is not yet available.

In this work, we design a universal cache-assisted MISO system capable of handling any instantaneous user distribution among caching profiles. Our system operation is comprised of two phases: \textit{i) Content placement phase}, and \textit{ii) content delivery phase}. In the placement phase, the server determines the number of caching profiles according to the cache ratio $\gamma$. Then, upon connecting to the network, each user is assigned to a single profile and stores the cache content of that profile.  During the content delivery phase, the server employs a clever combination of multicast and unicast transmissions to maximize the DoF. In this paper, we obtain closed-form expressions for the DoF, revealing the DoF loss caused by non-uniformness in users' distribution. Particularly, for the uniform user associations, it is shown that our proposed scheme achieves the optimal DoF not only in the regions covered in the literature but also in the region $\alpha \geq \hat{\eta}$ with non-integer $\nicefrac{\alpha}{\hat{\eta}}$.
Notably, our proposed scheme supports any user distribution, including the regions omitted in the existing literature~\cite{parrinello2020extending,parrinello2019fundamental,Abolpour2022CodedNetworks} such as networks with: \textit{i)} uneven user association with $\alpha> \hat{\eta}$ and non-integer $\frac{\alpha}{\hat{\eta}}$ unlike~\cite{parrinello2020extending}, \textit{ii)} $\min_{p} \eta_{p} \leq \alpha \leq \hat{\eta}$ unlike~\cite{parrinello2019fundamental}, and \textit{iii)} $\alpha \geq \hat{\eta}$ with $\alpha < K\gamma$ unlike~\cite{Abolpour2022CodedNetworks}.

 In this paper, bold lower-case and calligraphic letters show vectors and sets, respectively. $\left[ a:b \right]$ shows the set $\lbrace a,\cdots,b \rbrace$, $[a]=\lbrace 1,\cdots,a \rbrace$, $\left\vert \CA \right\vert$ is the cardinality of $\CA$, and for $\Lambda \subseteq \CA$, $\CA_{\backslash \Lambda}$ represents $\CA-\Lambda$.  $\left( \CA \Vert \CA \right)_{x}$ denotes $x$ concatenations of $\CA$ with itself, and $\CA \Vert \CB$ is the concatenation of $\CA$ and $\CB$.

\section{System Model}
In this paper, we focus on a dynamic MISO network, where a base station (BS) equipped with $L$ transmit antennas and   the spatial multiplexing gain of $\alpha \leq L$ serves several cache-enabled single-antenna users. The BS has access to a library $\CF$ with $N$ equal-sized files, and  each user is equipped with a large enough memory to store a portion $0< \gamma < 1$ of the entire library. We suppose $\gamma=\frac{\Brt}{P}$, where $\Brt$ and $P$ are natural numbers and $\gcd\left( \Brt, P \right)=1$. In this dynamic setup, users can move, enter and depart the network at any time. Accordingly, the BS does not have any prior knowledge about the number of available users during the transmission. When a user $k$ enters the network, it is assigned to a profile represented by $\Sfp [k] \in \left[ P \right]$, and the content of its cache is updated based on a \textit{content placement algorithm}.

In the placement phase, by following the same way as in \cite{maddah2014fundamental}, each file  $W^{n} \in \CF$, $n \in [N]$,  is split into $\binom{P}{\Brt}$ equal-sized mini-files $W^{n}_{\CP}$ such that $W^{n} \rightarrow \left\lbrace W^{n}_{\CP}: \CP \subseteq [P] ,  \left\vert \CP \right\vert=\Brt  \right\rbrace$.
 The cache content associated with profile $p \in [P]$, represented by $\CZ_{p}$, includes a portion $\gamma$  of each file $W^{n}$ as 
\begin{equation*}
    \CZ_{p}=\left\lbrace W^{n}_{\CP}: \CP \ni p, \CP \subseteq \left[ P \right], \left\vert \CP \right\vert = \Brt, \forall n \in \left[ N \right] \right\rbrace.
\end{equation*}
Then, defining $\CU_{p}$ as the set of users assigned to profile $p$, i.e., $\CU_{p}=\left\lbrace k: \Sfp \left[k \right]=p \right\rbrace$, each user $k \in \CU_{p}$ stores the cache content $\CZ_{p}$ during the  placement phase.

The dynamic nature of the network causes a fluctuating user population throughout time.  During regular intervals, the network's demanding users reveal their required files from the library $\CF$ to the BS. Then, using a \textit{content delivery algorithm}, the BS constructs and transmits a set of codewords, enabling users to retrieve their requested files. In this paper, we focus on the content delivery procedure over a specific time interval, where it is assumed that the number of present users during the BS's transmission is $K$.
In line with the general approach in the literature, we utilize the total DoF as the metric of interest, representing the average number of concurrent users served in parallel across all transmit intervals. 
The main contribution of this paper is to design delivery algorithms that provide a maximum combination of global caching and spatial multiplexing gains under the proposed dynamic conditions. As part of the delivery process, we discuss the transmission strategies to  reduce the DoF loss caused by non-uniformness in user-to-profile association in the following section. 
\begin{figure}[t]
      \centering
      \includegraphics[scale=0.25]{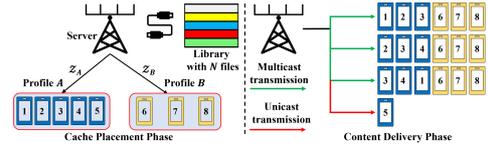}
      \caption{System model for a dynamic coded caching setup, where $P=2$, $\gamma=\frac{1}{2}$, $\Brt=1$, $\alpha=4$, $\hat{\eta}=4$, $Q=2$ and $\beta=3$. During the placement phase, each user assigned to profiles $A$ and $B$ stores the cache content associated with those profiles. For the delivery phase, user $5$ is served via unicasting and other users are served via $3$ multicast transmissions.}
      \label{fig:system}
      \vspace{-1em}
  \end{figure}

\section{Resource Allocation and Data Delivery}
\label{section: data delivery}
In this section, we discuss the resource allocation and transmission strategies during the content delivery phase. This phase commences once the set of active users reveals their requested files and comprises two consecutive steps: \textit{1) Coded caching (CC) data delivery}; and \textit{2) Unicast (UC) data delivery}. 

Let us define the number of users assigned to profile $p$ as the length of profile $p$ denoted by $\eta_{p}$, where without loss of generality, it is assumed that $\eta_{1} \geq \eta_{2} \geq  \cdots \geq \eta_{P}$. By choosing a \textit{delivery parameter}  $\hat{\eta}\leq \max_{p}\eta_{p}$\footnote{Here, the delivery parameter plays a similar role as the unifying length parameter in \cite{Abolpour2022CodedNetworks}. Both delivery and unifying length parameters tune the DoF loss caused by the non-uniformness in the user-to-profile association.}, the BS builds and transmits a set of codewords to serve at most $\hat\eta$ users assigned to each profile with a novel CC-based approach. In this regard, for every profile $p$
\\$\bullet$ if $\hat{\eta} < \eta_{p}$, we exclude $\eta_{p}-\hat{\eta}$ users, and exempt BS to serve these users during the CC delivery step. Accordingly, the excluded users are served in the UC delivery step.  
 \\$\bullet$ if $\hat{\eta} \geq \eta_{p}$, all $\eta_{p}$ users are served via the CC delivery step. 
 
 Now, let us suppose that the set of  users assigned to profile $p \in [P]$ and served during the CC delivery step is denoted by $\CV_{p}$ such that $\left\vert \CV_{p} \right\vert=\delta_{p}$, $\CV_{p}=\left\lbrace v_{p,1},v_{p,2}\cdots, v_{p,\delta_{p}} \right\rbrace$, and $v_{p,i} \in \CU_{p}$ for $i \in [\delta_{p}]$. We note that $\delta_{p}=\min \left( \hat{\eta},\eta_{p} \right)$, and clearly, $\delta_{1}=\hat{\eta}$ and $\delta_{1} \geq \delta_{2} \geq \cdots \geq \delta_{P}$.
 
In order to build the transmission vectors, as depicted in Fig.~\ref{fig:system}, the BS selects a parameter $Q$, which represents the number of profiles served in each transmission, and a parameter $\beta$, which expresses the number of users chosen from each profile to serve in each transmission. The necessary conditions for choosing any arbitrary values for $Q$ and $\beta$ are defined in Remark~\ref{rem: conditions}.
\begin{rem}
\label{rem: conditions}
In order to serve $Q$ profiles each with a maximum of $\beta$ users, the network parameters should satisfy the constraints $\Brt+1 \leq Q \leq \Brt+ \left\lceil \nicefrac{\alpha}{\beta} \right\rceil$ and $\beta \leq \min \left( \alpha, \hat{\eta} \right)$.
\end{rem}
\begin{IEEEproof}
    The proof is relegated to Appendix~\ref{apx: proof conditions}.
\end{IEEEproof}

In the following, we present the system operation maximizing the DoF performance separately for two regimes: \textit{i) $\alpha \leq  \hat{\eta}$} and \textit{ii) $\alpha > \hat{\eta}$}. For the CC delivery step, each of these cases operates either with \textit{Strategy A} (cf. Section~\ref{subsection: Strategy A}) or \textit{Strategy B} (cf. Section~\ref{subsection: strategy B}). The chosen transmission strategy only depends on the parameters $\alpha$ and $\hat{\eta}$, and it is independent of the content placement phase. In other words, the server performs the placement phase only based on parameter $\gamma$ without considering which transmission strategy the system will use for the CC delivery step.

\subsubsection{System operation for $\alpha \leq \hat{\eta}$}
\label{subsection: alpha l eta}
For the case of $\alpha \leq \hat{\eta}$, we set $\beta=\alpha$, and the only option for $Q$ to maximize the DoF is $Q=\Brt+1$. Here, \textit{Strategy A} is utilized to build the transmission vectors. Replacing the constraint $\alpha \leq \hat{\eta}$ with $\alpha \leq  \min_{p} \delta_{p}$ reduces our system model to \cite{parrinello2019fundamental}, while our proposed scheme also works for the scenarios with $\min_{p} \delta_{p} < \alpha \le \hat{\eta}$. 
\subsubsection{System operation for $\alpha > \hat{\eta}$}
\label{subsection: alpha g eta}
For this case, we set $\beta=\hat{\eta}$, and define $\hat{\alpha}=\frac{\alpha}{\hat{\eta}}$. If $\hat{\alpha}$ is an integer, then we follow \textit{Strategy~A} to build the transmission vectors. For non-integer $\frac{\alpha}{\hat{\eta}}$, the server can serve users via \textit{Strategy~A} by setting $\Brt+1 \leq Q \leq \Brt + \left\lfloor  \nicefrac{\alpha}{\hat{\eta}} \right\rfloor$, and via \textit{Strategy B} by choosing $Q=\Brt +\left\lceil \nicefrac{\alpha}{\hat{\eta}} \right\rceil$. In this case, if we assume that users are uniformly distributed among caching profiles, and  $\frac{\alpha}{\hat{\eta}}$ is an integer, the system performance is simplified to \cite{parrinello2020extending}, while our proposed scheme also covers the uneven user-to-profile associations with non-integer $\frac{\alpha}{\hat{\eta}}$.
\subsection{Transmission Strategy A}
\label{subsection: Strategy A}
In this strategy, each mini-file $W_{\CP}^{n}$  is split into $\beta \binom{P-\Brt -1}{Q-\Brt-1}$ subpackets $ W_{\CP ,q }^{n}$,  where $q \in [\beta \binom{P-\Brt -1}{Q-\Brt-1}]$ increases sequentially after each transmission to ensure none of subpackets is transmitted twice. Next, we follow the so-called \textit{elevation process} to serve $Q$ caching profiles each with at most $\beta$ users. 

\subsubsection*{Elevation process}
In this process, the aim is to characterize the set of users that are served in each transmission. Accordingly, we let $\phi_{p} = \max \left( \beta, \delta_{p} \right)$, and use the so-called \textit{transmission triple} $\left( r,c,l \right)$, where $r \in \left[ P-Q+1 \right]$, $c \in \left[ \phi_{r} \right]$ and $l \in \left[ \binom{P-r}{Q-1} \right]$. Also, it is assumed that $\eta_{1} \geq \eta_{2} \geq \cdots \geq \eta_{P}$, which results in $\phi_{1}\geq \phi_{2}\geq \cdots \geq \phi_{P}$. This process creates $\CT_{r,c,l}$, the set of users that are served during the transmission triple $\left(r,c,l \right)$. In this regard, first, for every $p \in [P]$, we elevate the set $\CV_{p}$ to the set $\CR_{p}$ as follows.
\begin{equation}
\label{eq: Rp final}
    \CR_{p}=\CR_{p,1} \Vert \cdots \Vert \CR_{p,\phi_{p}},
\end{equation}
where, for $j \in \left[  \phi_{p} \right]$, $\CR_{p,j}$ is defined as:
\begin{equation*}
\label{eq: Rpj}
\CR_{p,j}=
    \begin{cases}
      \CV_{p} & \delta_{p} \leq  {\beta} \\
       \left\lbrace v_{p,l}: l=\left( i+j -1 \right) \% \delta_{p}, 1\leq i \leq \beta \right\rbrace & \delta_{p} > \beta
    \end{cases}.
\end{equation*}
Here, $\%$ sign demonstrates the mod operator, for which $c \% c= c$ and $\left(d+c \right) \% c =d\%c$. Furthermore, we use generalized multiset definition where we allow the same elements to be repeated in the sets, e.g., $\left\lbrace a, a \right\rbrace$ cannot be reduced to $\left\lbrace a \right\rbrace$.  Now, for $p \in [P]$,  we define  $\CS_{p}=\CS_{p,1} \Vert \cdots \Vert \CS_{p, \hat{\eta}}$, while
  \begin{equation}
  \label{eq: Spj}
  \CS_{p,j}=
      \begin{cases}
        \CR_{p,j} & 1\leq j \leq \phi_{p}\\
       \varnothing &\phi_{p}+1 \leq j \leq \hat{\eta}
      \end{cases}.
  \end{equation}
 Then, for each $r \in [P-Q+1]$, we define the set
  $\CM_{r}$ as:
  \begin{equation}
  \label{eq: Mfinal}
      \begin{aligned}
            \CM_{r}=\left\lbrace \CF: \CF \subseteq \left[  r+1: P \right], \,  \left\vert \CF \right\vert =Q-1  \right\rbrace.
      \end{aligned}
  \end{equation}
 In the proceeding, we use $\CM_{r} \left( l \right)$ to indicate the $l$-th $\left( Q-1 \right)$-tuple of $\CM_{r}$. Finally,  the set $\CT_{r,c,l}$ for the transmission triple $\left( r,c,l \right)$  is given by: 
  \begin{equation}
  \small
  \label{eq: Tfinal}
      \begin{aligned}
           & \CT_{r,c,l}=
            \left\lbrace \CS_{r,c} \Vert \CS_{b_{1},c}  \Vert \cdots \Vert \CS_{b_{Q-1},c} : b_{i} \in  \CM_{r} \left( l \right), \forall i \in [Q-1]   \right\rbrace. 
      \end{aligned}
  \end{equation}
 Generally speaking, for the transmission triple $\left( r,c,l \right)$, if $\delta_{r}=0$, the BS does not transmit any signal; otherwise, the BS serves users assigned to  $\CT_{r,c,l}=\CS_{r,c} \Vert \CS_{b_{1},c}\Vert \cdots \Vert \CS_{b_{Q-1},c}$, while $\left\lbrace b_{1},\cdots , b_{Q-1} \right\rbrace=\CM_{r} \left( l \right)$. During the transmission triple $\left( r, c,l \right)$, by defining $\CN=\left\lbrace r \right\rbrace \cup \CM_{r} \left( l \right)$, the BS constructs the transmission vector as follows. 
  \begin{equation*}
      \begin{aligned}
            \Bx_{r,c,l} = \sum\limits_{\Lambda \subseteq \CN: \left\vert \Lambda \right\vert=\Brt} \, \,  \sum\limits_{k \in \CS_{p,c}: p \in \CN _{\backslash \Lambda}}  W_{\Lambda,q}^{k} \Bw_{\CG_{\Lambda}^{k}},
      \end{aligned}
  \end{equation*}
  where $\Bw_{\CG_{\Lambda}^{k}} \in \mathbb{C}^{L\times 1}$ is the zero-forcing (ZF) precoder that cancels out the interference of user $k$ at the set $\CG_{\Lambda}^{k}=\left\lbrace j \in \CS_{p,c}: \forall p \in \CN_{\backslash \Lambda}, j\neq k \right\rbrace$. In Appendix~\ref{apx: proof DoF}, it is proven that all users served with \textit{Strategy A} can decode their requested files at the end of the CC delivery step. Now,  assuming $\Bh_{i} \in \mathbb{C}^{L\times 1}$ as the channel gain of user $i$, at the end of the transmission triple $\left( r,c,l \right)$, the received signal at user $i$ is given by: 
  \begin{equation*}
      \begin{aligned}
            y_{i}=  \Bh_{i}^{\mathrm{H}}\Bx_{r,c,l}+ n_{i},
      \end{aligned}
  \end{equation*}
  where $n_{i}$ is the zero-mean additive white Gaussian noise of unit variance. In order to give further insight, an example for data delivery with \textit{Strategy A} is provided in Appendix~\ref{exmp: R}.

\subsection{Transmission Strategy B}
\label{subsection: strategy B}
When $\alpha > \hat{\eta}$ and $\frac{\alpha}{\hat{\eta}}$ is not an integer,  we can serve users by setting $\beta=\hat{\eta}$, and $Q=\Brt + \left\lceil \nicefrac{\alpha}{\hat{\eta}} \right\rceil$.\footnote{For $\alpha > \hat{\eta}$ and non-integer $\frac{\alpha}{\hat{\eta}}$, we can still serve users with \textit{Strategy A}. However, the achievable DoF is less than the one with \textit{Strategy B}.} Here, first, we split each mini-file into $\left( \hat{\eta}\Brt+ \alpha \right) \binom{P-\Brt-1}{Q-\Brt-1} \binom{Q-2}{Q-\Brt-2}$ subpackets. Then, we use the elevation process to serve $Q$ profiles each with at most $\beta$ users. 
\subsubsection*{Elevation process}
\label{subsection: alphageta nonint}
This process builds the set of users  served in each transmission.  First, for $r \in \left[ P \right]$,  we define $\CY_{r}$ as: 
\begin{equation}
    \begin{aligned}
           \CY_{r}=
           \begin{cases}
           \CV_{r} & \delta_{r}=\hat{\eta}\\
           \CV_{r} \Vert \left( f^{*} \Vert f^{*} \right)_{ \hat{\eta}-\delta_{r}} & \mathrm{o.w.}
           \end{cases},
    \end{aligned}
\end{equation}
where $f^{*}$ denotes the phantom (non-existent) users. Generally speaking, in each transmission of \textit{Strategy B}, we serve the users assigned to $Q$  profiles such that we select at most $\hat{\eta}$ users from $Q-1$ profiles, and pick $\theta=\alpha -\hat{\eta}\left\lfloor \nicefrac{\alpha}{\hat{\eta}} \right\rfloor$ users from another profile. Next, for $r \in \left[ P \right]$ and $m \in \left[ \hat{\eta} \right]$,  we consider the set $\CE_{r}^{m}$ as:
\begin{equation}
\label{eq: erm}
    \begin{aligned}
           \CE_{r}^{m}= \bigcup\nolimits_{i=0}^{\theta-1} \CY_{r} \left( \left( i+m \right) \% \hat{\eta} \right),
    \end{aligned}
\end{equation}
where $\CY_{r}(i)$ is the $i$-th element of $\CY_{r}$.  Indeed, $\CE_{r}^{m}$ shifts $\CY_{r}$ to the right for $m$ times, and picks $\theta$ elements from it. Then, for each $u \in \CE_{r}^{m}$, we define the set $\CK_{r}^{m,u}$ as follows. 
\begin{equation}
    \begin{aligned}
         \CK_{r}^{m,u}= \left( u  \Vert u \right)_{\nu_{1}} \Vert \left( f^{*}  \Vert f^{*} \right)_{\nu_{2}-\nu_{1}},
    \end{aligned}
\end{equation}
where $\nu_{1}=\binom{Q-2}{Q-\Brt-2}$ and $\nu_{2}=\binom{Q-1}{Q-\Brt-1}$. Next, by defining $\CK_{r}^{m,u} (s)$ as the $s$-th element of $\CK_{r}^{m,u}$, for $s \in \left[ \nu_{2} \right]$, it is assumed that $\CK_{r,s}^{m,u}$ is the $s$ circular shifts of $\CK_{r}^{m,u}$, such that:
\begin{equation}
\label{eq: Krsmu}
    \begin{aligned}
         \CK_{r,s}^{m,u}=\bigcup\nolimits_{i=0}^{\nu_{2}} \CK_{r}^{m,u} \left( \left( i+s \right) \% \nu_{2} \right).
    \end{aligned}
\end{equation}
Moreover, we assume that $\overline{\CP}_{r}=\left[ P \right]_{\backslash r}$ for $r \in \left[ P \right]$, and $\Bar{\delta}_{c}=\overline{\CP}_{r}(c)$ is the $c$-th element of $\overline{\CP}_{r}$. The caching profiles in $\overline{\CP}_{r}$ are sorted in descending order such that if $i<j$, then $\delta_{\overline{\CP}_{r}(i)} \geq \delta_{\overline{\CP}_{r}(j)}$. Next, for a given $r$ and $c \in \left[ P-Q+1 \right]$, let $\CI_{c}^{r}$ as:
\begin{equation}
\small
\label{eq: Icr non-int}
          \CI_{c}^{r}=
          \left\lbrace \CF: \CF \subseteq \left\lbrace \overline{\CP}_{r}(c+1),\cdots, \overline{\CP}_{r}(P-1) \right\rbrace, \left\lvert \CF \right\rvert=Q-2 \right\rbrace.
\end{equation}
Furthermore, denote the $l$-th $\left( Q-2 \right)$-tuple of $\CI_{c}^{r}$ by  $\CI_{c}^{r}(l)$, where $l \in \left[ \binom{P-c-1}{Q-2} \right]$. For the delivery process with \textit{Strategy~B}, we use the so-called \textit{transmission quintuple} $\left( r,c,l,m,s \right)$, where $r \in \left[ P \right]$, $c \in \left[ P-Q+1 \right]$, $l \in \left[ \binom{P-c-1}{Q-2} \right]$, $m \in \left[ \hat{\eta} \right]$ and $s \in \left[ \nu_{2} \right]$. In each transmission quintuple $\left( r,c,l,m,s \right)$, users assigned to the caching profiles $\CB=\left\lbrace \Bar{\delta}_{c} \right\rbrace \cup \CI_{c}^{r}(l)$, and users in the set $\CE_{r}^{m}$ are served. Suppose that $\CC=\left\lbrace \CF: \CF \subseteq \CB, \left\vert \CF \right\vert=\left\lfloor \nicefrac{\alpha}{\hat{\eta}} \right\rfloor \right\rbrace$, and $\CC(n)$ is the $n$-th $\left\lfloor \nicefrac{\alpha}{\hat{\eta}} \right\rfloor$-tuple of $\CC$ with $n \in \left[ \nu_{2} \right]$. 

We define the function $I^{+} \left( \Bar{\delta}_{c}, \CE_{r}^{m} \right)$ such that $I^{+} \left( \Bar{\delta}_{c}, \CE_{r}^{m} \right)=0$, if $\CE_{r}^{m}=f^{*}$ and $\Bar{\delta}_{c}=0$; otherwise, $I^{+} \left( \Bar{\delta}_{c}, \CE_{r}^{m} \right)=1$. 
 If $I^{+} \left( \Bar{\delta}_{c}, \CE_{r}^{m} \right)=1$, after eliminating the impacts of the phantom  users $f^{*}$, the BS builds the transmission vector for the transmission quintuple $\left( r,c,l,m,s \right)$ as follows. 
  \begin{equation*}
\label{eq: x non-int}
      \begin{aligned}
            \Bx_{r,c,l}^{m,s} = \sum\nolimits_{n=1}^{\nu_{2}} \, \, \sum\limits_{k \in  \CK_{r,s}^{m,u}(n) \cup \CV_{p}: u \in \CE_{r}^{m}, p \in \CC(n) }  W_{\Theta_{n},q}^{k} \Bw_{\CH_{\CC(n)}^{k}},
      \end{aligned}
  \end{equation*}
where $\Theta_{n}=\CB_{\backslash \CC(n)}$ with $\left\vert \Theta \right\vert=\Brt$, and $\Bw_{\CH_\CC(n)} \in \mathbb{C}^{L\times 1}$ is the precoder that suppresses the interference of user $k$ at the set $\CH_{\CC(n)}^{k}=\left\lbrace j \in \CE_{r}^{m} \cup \CV_{p}: p \in \CC(n),  j\neq k,f^{*} \right\rbrace$.  In Appendix~\ref{apx: proof DoF}, we prove that all users can decode their requested files with \textit{Strategy B}. Accordingly, user $i$ receives the signal
\begin{equation*}
      \begin{aligned}
            y_{i} =  \Bh_{i}^{\mathrm{H}}\Bx_{r,c,l}^{m,s}+ n_{i}.
      \end{aligned}
      \vspace{-.5em}
  \end{equation*}
 In order to give further insight, an example for data delivery with \textit{Strategy B} is provided in Appendix~\ref{exmp: Y}.


\subsection{Unicast (UC) Data Delivery}
\label{section: UC}
In the UC delivery step, the BS transmits data to the users excluded from  the CC delivery step. Here, unlike the CC delivery step that benefits from the global coded caching and spatial multiplexing gains, only local coded caching and spatial multiplexing gains are available. Suppose that the BS serves $K_{U}$ users during the UC delivery step such that $K_{U}=\sum_{p=1}^{P} \max \left( 0, \eta_{p}-\hat{\eta} \right)$. Each of the requested files by these users is split into the same number of subpackets as in the CC delivery step. Then, in order to transmit these missing subpackets, we follow a greedy algorithm similar to \cite{Abolpour2022CodedNetworks}, which comprises 3 processes:  1) sort  users based on the number of their missing subpackets in descending order; 2)~create a transmission vector to deliver one missing subpacket to each of the first $\min \left( \alpha, K_{U} \right)$ users; 3) repeat processes 1 and 2 until all missing files are transmitted.

\section{DoF Analysis}
In this section, we use DoF as the metric of interest to measure performance. Here, the DoF is defined as the  average number of users served concurrently during the delivery phase. In CC and UC delivery steps, we denote the total transmissions by $T_{M}$ and $T_{U}$, respectively, and the number of served users by $J_{M}$ and $J_{U}$. Therefore, DoF is computed as follows. 
\begin{equation}
\label{eq: DoF def}
    \mathrm{DoF}=\frac{J_{M}+J_{U}}{T_{M}+T_{U}}.
\end{equation}
Furthermore, we suppose that $K_{M}$ and $K_{U}$ users are served during the CC and UC delivery steps such that $K_{M}=\sum_{p} \min \left( \hat{\eta},\eta_{p} \right)$ and $K_{U}=\sum_{p} \max \left( 0, \eta_{p}-\hat{\eta} \right)$. The next theorem characterizes the DoF for the cache-aided networks operating with strategies \textit{A} and \textit{B} during the CC delivery step. 
\begin{thm}
\label{thm: DoF}
Consider a dynamic MISO network with the spatial multiplexing gain of $\alpha$, cache ratio $\gamma$ and the delivery parameter $\hat{\eta}$. If the system operates with \textit{Strategy A} in the CC delivery step, the DoF is given by:
    \begin{equation}
    \label{eq: DoF thm}
        \mathrm{DoF}=
        \begin{cases}
        \frac{K  \binom{P-1}{Q-1}\beta}{\sum\limits_{r=1}^{P-Q+1} D\left( \delta_{r}\right) \binom{P-r}{Q-1}} & K_{U}=0 \\
            \frac{K_{M} \binom{P-1}{Q-1}\beta + K_{U} \left( 1-\gamma  \right) \binom{P}{\Brt} \beta^{\prime}}{\sum\limits_{r=1}^{P-Q+1} D\left( \delta_{r}\right) \binom{P-r}{Q-1}+\left\lceil \frac{K_{U} \left( 1-\gamma \right) \binom{P}{\Brt}\beta^{\prime}}{\min \left( K_{U},\alpha \right)} \right\rceil} & K_{U} \neq 0
        \end{cases},
    \end{equation}
    where $\beta^{\prime}=\beta \binom{P-\Brt-1}{Q-\Brt-1}$ and $D(\delta_{r})= \phi_{r}$ if $\delta_{r} \neq 0$; otherwise, $D\left( \delta_{r} \right)=0$. If \textit{Strategy B} is applied during the CC delivery step, the DoF takes the form as follows.  
    \begin{equation}
    \label{eq: DoF thm non-int}
        \mathrm{DoF}=
        \begin{cases}
        \frac{K \binom{P-1}{Q-1}\left( \hat{\eta}\Brt+\alpha \right) \nu_{2}}{N_{M}} & K_{U}=0 \\
            \frac{K_{M} \binom{P-1}{Q-1}\left( \hat{\eta}\Brt+\alpha \right)\nu_{2} + K_{U} \left( 1-\gamma  \right) \binom{P}{\Brt} \alpha^{\prime}}{N_{M}+N_{U}} & K_{U} \neq 0
        \end{cases},
    \end{equation}
    where $\alpha^{\prime}=\left( \hat{\eta}\Brt+\alpha \right) \nu_{1} \binom{P-\Brt-1}{Q-\Brt-1}$, $N_{U}=\left\lceil \frac{K_{U} \left( 1-\gamma \right) \binom{P}{\Brt}\alpha^{\prime}}{\min \left( K_{U},\alpha \right)} \right\rceil$ and
    \begin{equation}
    \label{eq: NM def B}
        N_{M}=\sum_{r=1}^{P} \sum_{c=1}^{P-Q+1} \sum_{m=1}^{\hat{\eta}} \sum_{s=1}^{\nu_{2}} \binom{P-c-1}{Q-2}  I^{+} \left( \Bar{\delta}_{c}, \CE_{r}^{m} \right).
    \end{equation}
\end{thm}

\begin{IEEEproof}
The proof is relegated to Appendix~\ref{apx: proof DoF}.
\end{IEEEproof}
\begin{rem}
\label{rem: optimal}
Suppose that $K$ users are uniformly assigned to the caching profiles, i.e., $K=P\hat{\eta}$, and all users are served in the CC delivery step, i.e., $K_{M}=K$ and $K_{U}=0$. If $\alpha \leq \hat{\eta}$, our schemes achieves the optimal DoF $\alpha \left( P\gamma+1 \right)$ obtained in~\cite{parrinello2019fundamental}. If $\alpha > \hat{\eta}$,  the achievable DoF of our scheme is simplified to the optimal DoF $K\gamma+\alpha$ obtained in~\cite{parrinello2020extending}. However, unlike the scheme proposed in~\cite{parrinello2020extending}, our scheme supports the networks with non-integer $\frac{\alpha}{\hat{\eta}}$ (cf. Appendix~\ref{apx: proof DoF alphaleta uniform}). 
\end{rem}

Indeed, increasing non-uniformness in user distribution prevents the system from achieving optimal DoF performance. For instance, for the case $\alpha > \hat{\eta}$, suppose that all users are served with \textit{Strategy A} during the CC delivery step, such that $\beta=\hat{\eta}$ and $Q \leq \Brt + \left\lfloor \nicefrac{\alpha}{\hat{\eta}} \right\rfloor$. In this setup,  to boost the DoF performance, we can set $Q= \Brt+ \left\lfloor \nicefrac{\alpha}{\hat{\eta}} \right\rfloor$. By defining $\eta_{\rm avg}=\frac{K}{P}$ and assuming $\alpha$ is divisible by $\hat{\eta}$ and $\eta_{\rm avg}$, the DoF loss (compared to uniform user distribution) is  $\alpha \left( 1-\nicefrac{\eta_{\rm avg}}{\hat{\eta}} \right)$. 

Setting $Q=\Brt+ \left\lfloor \nicefrac{\alpha}{\hat{\eta}} \right\rfloor$, however,  requires to implement successive interference cancellation (SIC) at the receivers. To avoid using SIC, we can set $Q=\Brt +1$, which simplifies the DoF of the uniform association to $\eta_{\rm avg} \left( \Brt+1 \right)$. Here, we should compare the  achievable DoF of $\eta_{\rm avg} \left( \Brt+1 \right)$ with $\alpha$ such that: \textit{i)} if $\eta_{\rm avg} \left( \Brt+1 \right) \geq \alpha$, we use the proposed CC scheme to simultaneously benefit from global CC and spatial multiplexing gains; \textit{ii)} if $\eta_{\rm avg} \left( \Brt+1 \right)<\alpha$, we serve users via unicasting to not have any loss in  DoF. 





\begin{rem}
\label{rem: DoF max alphaleta}
For the non-uniform user-to-profile association with  $Q=\Brt +1$ and $\eta_{1} \leq \alpha$, the best possible DoF is achievable by setting $\hat{\eta}=\eta_{1}$ (cf. Appendix~\ref{apx: Proof DoF max alphaleta}).
\end{rem}





\section{Numerical Results}
In this section, we examine the impacts of non-uniformness in user distribution on the achievable DoF. In this regard, assume a cache-aided MISO setup with $\Brt=1$, $P=6$ and $\gamma=\frac{1}{6}$, in which  $K=30$ users are present during  the delivery phase. Here, for each association, we compute the standard deviation $\sigma$ as $\sigma^{2}=\frac{1}{P} \sum_{p=1}^{P} \left( \eta_{p} -\eta_{\rm avg} \right)^{2}$, where $\eta_{\rm avg}=5$. 

\begin{figure*}[!htb]
\centering
\subfloat[$\sigma=0$]{\includegraphics[scale=0.36]{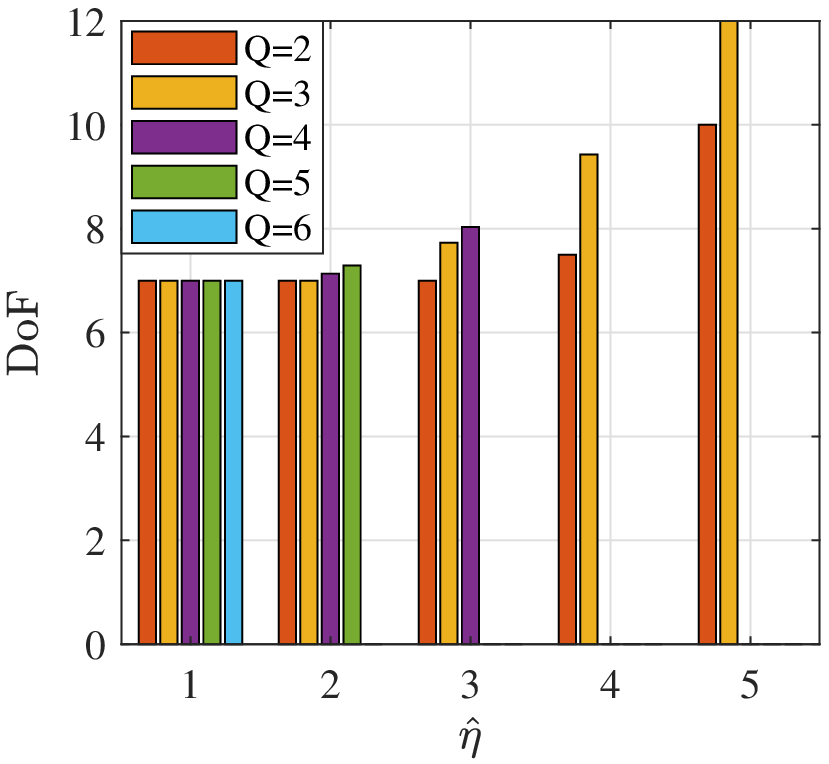}\label{fig: sigma0}} 
\subfloat[$\sigma=0.3$]{\includegraphics[scale=0.36]{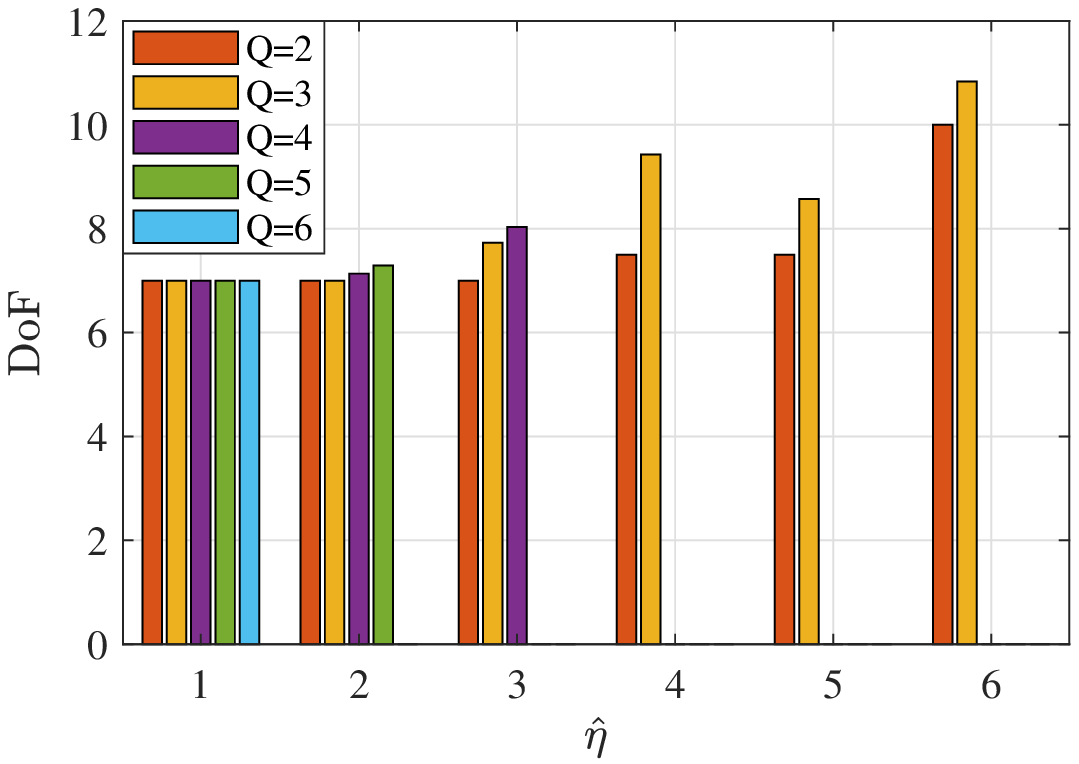} \label{fig: sigma03}} 
\subfloat[$\sigma=3$]{\includegraphics[scale=0.36]{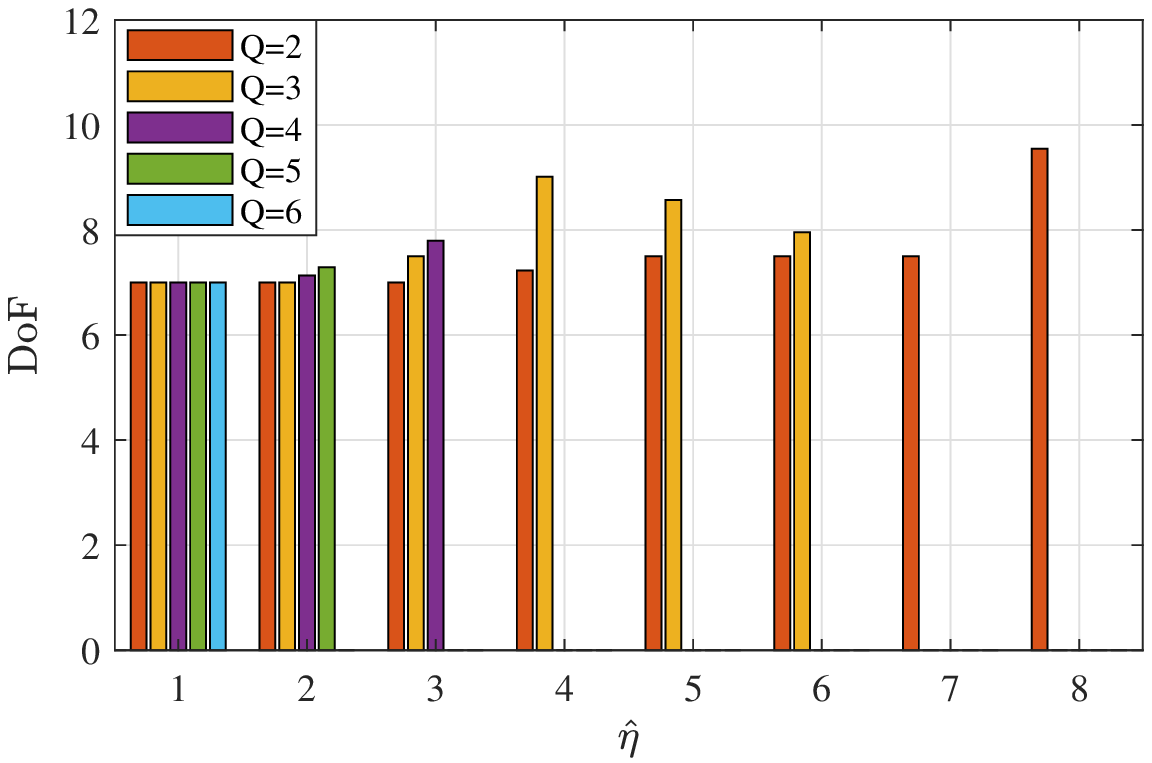}\label{fig: sigma3}}
\caption{The DoF versus $\hat{\eta}$ with different values of $Q$, $K=30$, $\alpha=7$, $P=6$, $\gamma=\frac{1}{6}$ and $\Brt=1$ for: (a) $\sigma=0$, (b) $\sigma=0.3$ and (c) $\sigma=3$.}
 \label{fig: DoF}
 \vspace{-1em}
\end{figure*}
Fig.~\ref{fig: DoF} illustrates the maximum achievable DoF for different $Q$ and $\sigma$ values with $\alpha=7$. As observed from Fig.~\ref{fig: sigma0}, for the uniform user distribution, i.e., $\sigma=0$, our scheme achieves the optimal DoF $K\gamma+\alpha=12$ with $\hat{\eta}=\eta_{\rm avg}=5$ and $Q=\Brt+\lceil \nicefrac{\alpha}{\hat{\eta}} \rceil=3$. Here, we note that the system performance during the CC delivery step corresponds to \textit{Strategy B} for the regime $\alpha> \hat{\eta}$ and non-integer $\nicefrac{\alpha}{\hat{\eta}}$, which is missing in the literature.  For small $\sigma$ values (e.g., $\sigma=0.3$),  although the achievable DoF for $\hat{\eta}=6$ and $Q=\Brt+1=2$  is slightly less than $\hat{\eta}=6$ and $Q=3$,   the receiver structure for $Q=2$ is more straightforward, as they do not need to implement SIC. For large  $\sigma$ values (e.g., $\sigma=3$), when $\max_{p} \eta_{p} > \alpha$, setting $Q=\Brt+1$ and $\hat{\eta}=\max_{p} \eta_{p}$  maximizes the achievable DoF.

\begin{figure}[!htb]
    \centering
    \includegraphics[scale=.4]{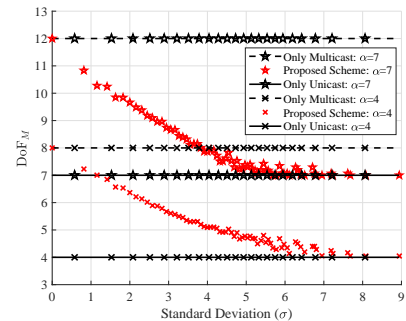}
    \caption{The average of the maximum achievable DoF $\left( {\rm DoF}_{M} \right)$ versus the standard deviation $(\sigma)$ with $K=30$, $\gamma=\frac{1}{6}$, $P=6$, $\Brt=1$ and $\alpha \in \left\lbrace 4,7 \right\rbrace$.}
    \label{fig: DoFM}
    \vspace{-.6em}
\end{figure}
In Fig.~\ref{fig: DoFM}, for any association, we find $\mathrm{DoF}_{\max}$ which is the maximum achievable DoF obtained by a line search over $\hat{\eta}$ and $Q$ values. Accordingly, for the associations with the same $\sigma$ value, $\mathrm{DoF}_{M}$ indicates the average of $\mathrm{DoF}_{\max}$. Here, our proposed scheme is compared with the optimal case, which corresponds to uniform user distribution (described as \textit{only multicast}), and the case that all users are served via unicasting (described as \textit{only unicast}). Although the placement phase was designed solely based on $\gamma$, our scheme boosts the maximum achievable DoF by $10\%-70\%$ over unicasting for moderate $\sigma$~(e.g., $\sigma=1-4.5$). So, to maximize the achievable DoF, the server should serve users via the proposed approach for moderate $\sigma$, and serve users via unicasting for large $\sigma$. 

\section{Conclusion}
We proposed a novel coded caching scheme for handling network dynamicity where the users can freely enter or depart the network at any time.
The conventional schemes in the literature are not truly dynamic as they are only applicable if:
\textit{1)} minimum profile length (the number of users assigned to the profile) exceeds the spatial multiplexing gain $\alpha$, and \textit{2)} $\alpha \geq \hat{\eta}$ and $\alpha$ exceeds the global CC gain, where $\hat{\eta}$ can be the length of any profile. 
Our proposed scheme 
addressed this bottleneck by providing a universal solution applicable to any dynamic network setup, removing all the constraints imposed by existing solutions.
We also analyzed the degrees-of-freedom (DoF) performance of the proposed scheme, and for the uniform distribution, we showed that it achieves the optimal DoF not only in the regions covered in the literature but also in the region $\alpha \geq \hat{\eta}$ with non-integer $\nicefrac{\alpha}{\hat{\eta}}$.




\appendices

\section{Proof of Remark~\ref{rem: conditions}}
\label{apx: proof conditions}
As stated earlier, each transmission vector is constructed to serve at most $Q\beta$ users. Hence, each of these $Q\beta$ users interfered by $Q\beta-1$ interfering users. Since each user has stored a $\gamma$ portion of the entire library, it can remove at most $\beta P \gamma=\beta \Brt$ interference terms using the stored cache content. Therefore, $\alpha$ must be large enough to suppress the remaining $Q\beta-1-\Brt \beta$ interference terms, i.e., 
\begin{equation}
    \alpha-1 \geq Q\beta-1-\Brt \beta,
\end{equation}
which results in $Q \leq \Brt + \left\lceil \frac{\alpha}{\beta} \right\rceil$. Furthermore, $Q\geq \Brt+1$ should hold to benefit from the maximum coded caching gain.  

Let us assume that profile $p$ is one of the $Q$ selected profiles for the transmission. Therefore, $\min \left( \beta, \delta_{p} \right)$ users are served from profile $p$ during this transmission. On the one hand, it is supposed that the maximum number of users of each profile served in the CC delivery step is $\hat{\eta}$, which results in having $\beta \leq \hat{\eta}$. On the other hand, for each profile $p \in \left[ P \right]$, we should have $\min \left( \beta, \delta_{p} \right) \leq \alpha$ to be able to cancel out the inter-stream interference between users assigned to the same profile. Hence, according to the fact that $\delta_{1}=\hat{\eta}$, one can say $\min \left( \beta, \delta_{1} \right)=\min \left(\beta, \hat{\eta} \right)=\beta \leq \alpha$, which results in having the necessary condition $\beta \leq \min \left( \alpha,\hat{\eta} \right)$ to select the parameter $\beta$.

\section{Example for \textit{Strategy A}}
\label{exmp: R}
For a cache-aided dynamic setup with $P=3$, $\gamma=\frac{1}{3}$, $\Brt=1$, $\alpha=6$, $\beta=3$ and $Q=\Brt+\left\lfloor \nicefrac{\alpha}{\beta} \right\rfloor= 3$, let us assume that $\CU_{1}=\left\lbrace 1,2,3,4,5 \right\rbrace$, $\CU_{2}=\left\lbrace 6,7,8,9 \right\rbrace$ and $\CU_{3}=\left\lbrace 10,11,12 \right\rbrace$ are the set of users assigned to profiles $1,2$ and $3$, respectively. 
During the placement phase, each file $W^{n}$ is split into $\binom{P}{\Brt}=3$ mini-files $W_{\CP}^{n}$, where $\CP \in \left\lbrace 1,2,3 \right\rbrace$. For $n \in \left[ N \right]$, users assigned to profiles $1,2$ and $3$ store the mini-files $W_{1}^{n}$, $W_{2}^{n}$ and $W_{3}^{n}$, respectively.  For the delivery phase, each mini-file $W_{\CP}^{n}$ is further split into $\beta \binom{P-\Brt-1}{Q-\Brt-1}=3$ subpackets $W_{\CP,q}^{n}$, where $q \in \left[ 3 \right]$. 

Assuming, $\hat{\eta}=4$, we have $\CV_{1}=\left\lbrace 1,2,3,4 \right\rbrace$, $\CV_{2}=\left\lbrace 6,7,8,9 \right\rbrace$ and $\CV_{3}=\left\lbrace 10,11,12 \right\rbrace$, while user $5$ will be served during the UC delivery step\footnote{Although  it is mentioned in Section~\ref{section: data delivery} to set $\beta=\hat{\eta}$ to maximize the achievable DoF for the case of $\alpha>\hat{\eta}$, here, we set $\beta<\hat{\eta}$ to give further insight on the system operation with \textit{Strategy A}.}. Then, we obtain $\CR_{1}-\CR_{3}$ as follows. 
\begin{equation*}
\begin{aligned}
    &\CR_{1}=\left(  \left\lbrace 1,2,3 \right\rbrace  \Vert \left\lbrace 2,3,4 \right\rbrace  \Vert  \left\lbrace 3,4,1 \right\rbrace \Vert  \left\lbrace 4,1,2 \right\rbrace    \right),\\
     &\CR_{2}=\left(  \left\lbrace 6,7,8 \right\rbrace\Vert  \left\lbrace 7,8,9 \right\rbrace\Vert   \left\lbrace 8,9,6 \right\rbrace\Vert   \left\lbrace 9,6,7 \right\rbrace  \right), \\
     & \CR_{3}=\left( \left\lbrace 10,11,12 \right\rbrace \Vert  \left\lbrace 10,11,12 \right\rbrace\Vert  \left\lbrace 10,11,12 \right\rbrace \right),\\
    \end{aligned}
    \end{equation*}
where $\CR_{1,1}=\left\lbrace 1,2,3 \right\rbrace$ and so on. Accordingly, using $\CR_{1}-\CR_{3}$, $\CS_{1}-\CS_{3}$ are given by:
 \begin{equation*}
     \begin{aligned}
           &\CS_{1}=\left(  \left\lbrace 1,2,3 \right\rbrace  \Vert \left\lbrace 2,3,4 \right\rbrace  \Vert  \left\lbrace 3,4,1 \right\rbrace \Vert  \left\lbrace 4,1,2 \right\rbrace    \right),\\
           &\CS_{2}=\left(  \left\lbrace 6,7,8 \right\rbrace\Vert  \left\lbrace 7,8,9 \right\rbrace\Vert   \left\lbrace 8,9,6 \right\rbrace\Vert   \left\lbrace 9,6,7 \right\rbrace  \right), \\
           & \CS_{3}=\left( \left\lbrace 10,11,12 \right\rbrace \Vert  \left\lbrace 10,11,12 \right\rbrace\Vert  \left\lbrace 10,11,12 \right\rbrace \Vert \varnothing \right).
     \end{aligned}
 \end{equation*}
In order to build the transmission vector for the transmission triple $\left( 1,1,1 \right)$, first, we obtain $\CM_{1}=\CM_{1}(1)= \left\lbrace 2,3 \right\rbrace$. Therefore, the set of users that are served during the transmission triple $(1,1,1)$ is equal to:
\begin{equation*}
          \begin{aligned}
                \CT_{1,1,1} &=
                \left( \CS_{1,1} \Vert \CS_{2,1} \Vert \CS_{3,1}\right)
               = \left\lbrace 1,2,3,6,7,8,10,11,12 \right\rbrace.
          \end{aligned}
      \end{equation*}
      Finally, the transmission vector corresponding to the  transmission triple $\left( 1,1,1 \right)$ is constructed as follows. 
  \begin{equation*}
      \begin{aligned}
           \Bx_{1,1,1}= & \sum\limits_{\Lambda \subseteq \left\lbrace 1,2,3 \right\rbrace: \left\vert \Lambda \right\vert=1} \, \, \sum\limits_{k \in \CS_{p,c}: p \in \left\lbrace 1,2,3 \right\rbrace_{\backslash \Lambda}}  W_{\Lambda,q}^{k} \Bw_{\CG_{\Lambda}^{k}} \\
           =  &\sum\limits_{k \in \CS_{2,1} \Vert \CS_{3,1}}  W_{1,q}^{k} \Bw_{\CG_{1}^{k}} +
            \sum\limits_{k \in \CS_{1,1} \Vert \CS_{3,1}}  W_{2,q}^{k} \Bw_{\CG_{2}^{k}} \\
          + & \sum\limits_{k \in \CS_{1,1} \Vert \CS_{2,1}}  W_{3,q}^{k} \Bw_{\CG_{3}^{k}}.
      \end{aligned}
  \end{equation*}
 Now, in order to show the decoding process, for any $k,j \in \CS_{p,c}$ with $p \in \CN_{\backslash \Lambda}$, if $j\neq k$, using ZF precoders, we have $\Bh_{k}^{\rm H} \Bw_{\CG_{\Lambda}^{j}}=0$; otherwise ($j=k$), $\Bh_{k}^{\rm H} \Bw_{\CG_{\Lambda}^{k}}=1$. For example, $\Bh_{1}^{\rm H} \Bw_{\CG_{3}^{1}}=1$ and $\Bh_{2}^{\rm H} \Bw_{\CG_{3}^{1}}=0$. Now, the received  signal at user $1$ is given by:
  \begin{equation*}
      \begin{aligned}
            y_{1}= \sum\limits_{k \in \CS_{2,1} \Vert \CS_{3,1}} \Bh_{1}^{\rm H}  \Bw_{\CG_{1}^{k}}W_{1,q}^{k}+W_{2,1}^{1}+W_{3,1}^{1}+n_{1}.
      \end{aligned}
  \end{equation*}
  User $1$ have all subpackets $W_{1,q}^{n}$, $\forall n \in [N]$ in its cache memory, therefore, it can regenerate $\sum\limits_{k \in \CS_{2,1} \Vert \CS_{3,1}} \Bh_{1}^{\rm H}  \Bw_{\CG_{1}^{k}}W_{1,q}^{k}$ and subtract it from $y_{1}$, which results in observing the superposition signal $\Tilde{y}_{1}=W_{2,1}^{1}+W_{3,1}^{1}+n_{1}$. Here, we note that user $1$ needs to implement SIC to decode the subpackets~\cite{tolli2017multi},~\cite{parrinello2019fundamental}. 

\section{Example for \textit{Strategy B}}
  \label{exmp: Y}
With the given $\CU_{1}-\CU_{3}$ and $\CV_{1}-\CV_{3}$ in Appendix~\ref{exmp: R}, suppose that $P=3$, $\gamma=\frac{1}{3}$, $\Brt=1$, $\alpha=6$, $\beta=\hat{\eta}=4$, $\theta=\alpha -\hat{\eta}\left\lfloor \nicefrac{\alpha}{\hat{\eta}} \right\rfloor=2$ and $Q=\Brt+\left\lceil \nicefrac{\alpha}{\hat{\eta}} \right\rceil=3$. For the placement phase, similarly to Appendix~\ref{exmp: R}, each file $W^{n}$ is split into 3 mini-files, and users assigned to profiles $1,2$ and $3$, store the mini-files $W_{1}^{n}$, $W_{2}^{n}$ and $W_{3}^{n}$, respectively. Then, during the content delivery phase, each mini-file is split into  $\left( \hat{\eta}\Brt+ \alpha \right) \binom{P-\Brt-1}{Q-\Brt-1} \binom{Q-2}{Q-\Brt-2}=10$ subpackets. Now, for the transmission quintuple $(1,1,1,1,1)$, we have:
\begin{equation*}
    \begin{aligned}
           \CY_{1}=\left\lbrace 1,2,3,4 \right\rbrace,   \, \, \CY_{2}=\left\lbrace 6,7,8,9 \right\rbrace, \, \,   \CY_{3}=\left\lbrace 10,11,12,f^{*} \right\rbrace.
    \end{aligned}
\end{equation*}
At the next step, for $r=c=1$, it is found that $\overline{\CP}_{1}=\left\lbrace 2,3 \right\rbrace$,  $\overline{\CP}_{1}(1)= 2$, and $\CI_{1}^{1}=\CI_{1}^{1}(1)=\left\lbrace 3 \right\rbrace$. During this transmission, the users of the set $\CE_{1}^{1}=\left\lbrace 1,2 \right\rbrace$, and the users assigned to the profiles $\CB=\overline{\CP}_{1}(1)\cup \CI_{1}^{1}(1)=\left\lbrace 2,3 \right\rbrace $ are served. Furthermore, $\CK_{1,1}^{1,1}=\left\lbrace 1,f^{*} \right\rbrace$ and $\CK_{1,1}^{1,2}=\left\lbrace 2,f^{*} \right\rbrace$. In addition, it is found that $\CC=\left\lbrace  \left\lbrace  2  \right\rbrace, \left\lbrace 3 \right\rbrace \right\rbrace$ such that $\CC(1)=\left\lbrace 2 \right\rbrace$ and $\CC(2)=\left\lbrace 3 \right\rbrace$. Accordingly, we have $\Theta_{1}=3$ and $\Theta_{2}=2$.

Hence, after removing the phantom users $f^{*}$, the transmission vector for the transmission quintuple $(1,1,1,1,1)$ takes the form of:
\begin{equation*}
      \begin{aligned}
            \Bx_{1,1,1}^{1,1} &= \sum\limits_{n=1}^{2} \, \sum\limits_{k \in  \CK_{r,s}^{m,u}(n) \cup \CV_{p}: u \in \left\lbrace 1,2 \right\rbrace, p \in \CC(n) }  W_{\Theta_{n},q}^{k} \Bw_{\CH_{\CC(n)}^{k}}\\
             &=\sum\limits_{k \in \left\lbrace 1,2,6,7,8,9 \right\rbrace}  W_{3,q}^{k} \Bw_{\CH_{2}^{k}}+\sum\limits_{k \in \left\lbrace 10,11,12 \right\rbrace}  W_{2,q}^{k} \Bw_{\CH_{3}^{k}},
      \end{aligned}
  \end{equation*}
  where  $\CH_{2}^{6}=\left\lbrace 1,2,7,8,9 \right\rbrace$, $\CH_{3}^{10}=\left\lbrace 1,2,11,12 \right\rbrace$ and so on. Similarly to Appendix~\ref{exmp: R}, all users that are served via \textit{Strategy~B} are able to decode their requested files.

\section{Proof of Theorem~\ref{thm: DoF}}
\label{apx: proof DoF}
Here, first, we present a lemma to simplify the calculations, then, prove the theorem for strategies \textit{A} and \textit{B} in Sections~\ref{subsection: proof A} and \ref{subsection: proof B}, respectively.
\begin{lem}
    \label{lemma: comb}
    Given $P$ and $Q$, we have $\sum\limits_{r=1}^{P-Q+1} \binom{P-r}{Q-1}=\binom{P}{Q}$.
    \end{lem}
    
    \begin{IEEEproof}
        Assume a scenario at which we have a set $\CX=\left\lbrace b_{1}, \cdots, b_{P}  \right\rbrace$ with $\left\vert \CX \right\vert =P$, and we want to select the subsets $\Bar{\CX} \subseteq \CX$ such that $\left\vert \Bar{\CX} \right\vert=Q$. Here, our aim is to count the possibilities for the subsets $\Bar{\CX}$. One way to count these possibilities is to simply pick the subsets $\Bar{\CX}$ from $\CX$, and then count them. Here,  clearly, the number of possibilities is equal to $\binom{P}{Q}$. Another way to choose $\Bar{\CX}$ is to define $\left(P-Q+1 \right)$ subsets $\CL_{r}$ first, such that $r \in \left[ P-Q+1 \right]$, $\CL_{r}=\left\lbrace  b_{i}: r+1 \leq i \leq P \right\rbrace$ and $\left\vert \CL_{r} \right\vert=P-r$. Then, the selection of $\Bar{\CX}$ consists of $\left(P-Q+1 \right)$ rounds. In round $r \in \left[ P-Q+1 \right]$, we pick $b_{r}$ and choose $\left( Q-1 \right)$ remaining elements from $\CL_{r}$. In this regard, the number of possibilities to choose $\Bar{\CX}$ in round $r$ is equal to $\binom{P-r}{Q-1}$. By following this approach, we have $\sum\limits_{r=1}^{P-Q+1} \binom{P-r}{Q-1}$ possibilities to choose $\Bar{\CX}$. Hence, it is sure that $\sum\limits_{r=1}^{P-Q+1} \binom{P-r}{Q-1}=\binom{P}{Q}$.
    \end{IEEEproof}

\subsection{Proof of Theorem~\ref{thm: DoF} for \textit{Strategy A}}
\label{subsection: proof A}
In order to prove the theorem, first, we compute the number of times that each user is served with \textit{Strategy A} during the CC delivery step. Suppose that user $k$ with $p\left[k \right]=p$ is present in the CC delivery step. As per \eqref{eq: Mfinal} and \eqref{eq: Tfinal}, user $k$ is served in the  transmission triple $\left( r,c,l \right)$ such that $r \in \left[ p \right]$. On the other hand, according to \eqref{eq: Rp final} and \eqref{eq: Spj}, for each $r \in [p]$, there are $\beta$ values for $c$ that user $k$ is served in the transmission triple $\left( r,c,l \right)$. Now, for any $r \in [p]$, let $\CA_{k}^{(r)}$ be the set of values for $c$ at which user $k$ is served in the transmission triple $\left( r,c,l \right)$ such that $\left\vert \CA_{k}^{(r)} \right\vert=\beta$. As mentioned in Section~\ref{subsection: Strategy A}, for $r=p$ and $c \in  \CA_{k}^{(p)} $, there are $ \binom{P-p}{Q-1}$ values for $l$ at which user $k$ is served during the transmission triple $\left( r,c,l \right)$. Furthermore, for $r \in \left[ p-1 \right]$ and $c \in  \CA_{k}^{(r)}$, there exist $\binom{P-r-1}{Q-2}$  transmission triples  in which user $k$ receives the subpackets. Therefore, the number of times that user $k$ is served with \textit{Strategy A} during the CC delivery step is given by:
  \begin{equation}
  \label{eq: gpk def}
      \begin{aligned}
        \beta \left( \binom{P-p}{Q-1}+ \sum\limits_{r=1}^{p-1}\binom{P-r-1}{Q-2} \right).
      \end{aligned}
  \end{equation}
 In the proceeding, we simplify \eqref{eq: gpk def}. We know that $\binom{P-p}{Q-1}+\binom{P-p}{Q-2}=\binom{P-p+1}{Q-1}$, which results in having $\binom{P-p}{Q-1}+\binom{P-p}{Q-2}+\binom{P-p+1}{Q-2}=\binom{P-p+1}{Q-1}+\binom{P-p+1}{Q-2}=\binom{P-p+2}{Q-1}$. Therefore, by following a similar procedure, we can say:
 \begin{equation}
 \label{eq: gpk final}
     \begin{aligned}
         \binom{P-p}{Q-1}+ \sum\limits_{r=1}^{p-1}\binom{P-r-1}{Q-2}=  \binom{P-1}{Q-1}.
     \end{aligned}
 \end{equation}
 By substituting \eqref{eq: gpk final} into \eqref{eq: gpk def}, it is observed that $\beta \binom{P-1}{Q-1}$ transmissions serve user $k$. Now, since there are $K_{M}$ users in the CC delivery step, the total number of times that all users served with \textit{Strategy A} is equal to:
 \begin{equation}
 \label{eq: JM A apx}
     J_{M}=K_{M} \beta \binom{P-1}{Q-1}.
 \end{equation}
 
 In the proceeding, we show that all users can decode their requested files with \textit{Strategy A}. As per \eqref{eq: Rp final}-\eqref{eq: Tfinal},  user $k$ served with \textit{Strategy A} receives $\binom{Q-1}{Q-\Brt-1}$ subpackets in each  transmission. Hence, the total number of received subpackets at user $k$ is given by:
\begin{equation}
\label{eq: sub int}
    \begin{aligned}
        \beta \binom{P-1}{Q-1}\binom{Q-1}{Q-\Brt-1}=\left( 1-\gamma \right) \beta \binom{P}{\Brt} \binom{P-\Brt -1}{Q-\Brt-1}.
    \end{aligned}
\end{equation}
 On the other hand, each file is split into $\beta \binom{P}{\Brt} \binom{P-\Brt -1}{Q-\Brt-1}$ subpackets, while a $\gamma$ portion of these subpackets is stored in cache memory of user $k$ during the placement phase, and $\left( 1-\gamma \right) \beta \binom{P}{\Brt} \binom{P-\Brt -1}{Q-\Brt-1}$ missing subpackets must be delivered to user $k$  with \textit{Strategy A}. As observed from \eqref{eq: sub int}, user $k$  receives all $\left( 1-\gamma \right) \beta \binom{P}{\Brt} \binom{P-\Brt -1}{Q-\Brt-1}$ missing subpackets with \textit{Strategy~A}, hence, one can conclude that all users attending the CC delivery step with \textit{Strategy A} are able to decode their requested files. 
 
 Now, we compute the number of transmitted subpackets during the UC delivery step. By following a similar way as in~\cite{Abolpour2022CodedNetworks},   each user served with  UC delivery step must receive $\left( 1-\gamma \right) \beta \binom{P}{\Brt} \binom{P-\Brt -1}{Q-\Brt-1}$ subpackets. Since there are $K_{U}$ users in the UC delivery step, the number of transmitted subpackets during the UC delivery step is
 \begin{equation}
 \label{eq: JU A apx}
     J_{U}=K_{U}\left( 1-\gamma \right) \binom{P}{\Brt}\beta^{\prime}.
 \end{equation}
 Using \eqref{eq: JM A apx} and \eqref{eq: JU A apx}, the total transmitted subpackets in the data delivery phase is given by:
 \begin{equation}
 \label{eq: JM plus JU}
     \begin{aligned}
         J_{M}+J_{U}=K_{M} \binom{P-1}{Q-1} \beta+K_{U}\left( 1-\gamma \right) \binom{P}{\Brt}\beta^{\prime}.
     \end{aligned}
 \end{equation}
 
 Next, we calculate the total number of transmissions in the CC delivery step. As per \eqref{eq: Tfinal}, for fixed $r=1$ and $c=c_{0}\in \left[ \phi_{1} \right]$, users assigned to profiles $\CS_{1,c_{0}}$, $\CS_{b_{1},c_{0}}, \cdots , \CS_{b_{Q-1},c_{0}}$ are served  during the transmission triple $\left( 1,c_{0},l\right)$, where $\left\lbrace  b_{1}, \cdots, b_{Q-1} \right\rbrace=\CM_{1}\left( l \right)$ and $l \in \left[ \binom{P-1}{Q-1}  \right]$. Hence, the number of transmissions for $r=1$ is equal to $\phi_{1} \binom{P-1}{Q-1}$.  Assuming $\delta_{2}>0$, for $r=2$ and $c=c_{0} \in \left[ \phi_{2} \right]$, the users assigned to $\CS_{1,c_{0}}$ are not served in the  transmission triple $\left( 2,c_{0},l \right)$, and the BS serves users assigned to $\CS_{2,c_{0}}, \CS_{b_{1},c_{0}} \cdots , \CS_{b_{Q-1},c_{0}}$, such that $\left\lbrace b_{1}, b_{2}, \cdots, b_{Q-1} \right\rbrace=\CM_{2} \left(l \right)$, and $l \in \left[ \binom{P-2}{Q-1} \right]$. Therefore, the number of transmissions for $r=2$ is $\phi_{2} \binom{P-2}{Q-1}$. Similarly, assuming $\delta_{r} > 0$,  for any  $r \in \left[ P-Q+1 \right]$ and $c_{0} \in \left[ \phi_{r} \right]$, users assigned to the set $\CT_{r,c_{0},l}=\CS_{r,c_{0}} \Vert \CS_{b_{1},c_{0}} \Vert \cdots \Vert \CS_{b_{Q-1},c_{0}}$ are served, where $\left\lbrace b_{1},\cdots , b_{Q-1} \right\rbrace=\CM_{r} \left( l \right)$ and $l \in \left[ \binom{P-r}{Q-1} \right]$. So, supposing $\delta_{r}>0$, the total number of transmissions for any $r \in \left[ P-Q+1 \right]$ is equal to $\phi_{r} \binom{P-r}{Q-1}$.
 As a result, the total number of transmissions in the CC delivery step with \textit{Strategy~A}  is given by:
 \begin{equation}
 \label{eq: TM}
     \begin{aligned}
         T_{M}=\sum\limits_{r=1}^{P-Q+1} D\left( \delta_{r}\right) \binom{P-r}{Q-1}.
     \end{aligned}
 \end{equation}
 
Then, we obtain the total number of transmissions during the UC delivery step. As stated earlier, during this step, we have to deliver $K_{U} \left( 1-\gamma \right)  \binom{P}{\Brt} \beta^{\prime} $ subpackets via $T_{U}$ transmissions, such that in each transmission, $\min \left( K_{U},\alpha \right)$ subpackets are transmitted. So, given $K_{U} \neq 0$, the total number of transmissions to deliver these subpackets is equal to:
 \begin{equation}
 \label{eq: TU}
     \begin{aligned}
         T_{U}= \left\lceil \frac{K_{U} \left( 1-\gamma \right) \binom{P}{\Brt}\beta^{\prime}}{\min \left( K_{U},\alpha \right)} \right\rceil.
     \end{aligned}
 \end{equation}
 
 Now, for $K_{U} \neq 0$, by substituting \eqref{eq: JM plus JU}-\eqref{eq: TU} into \eqref{eq: DoF def}, the DoF is obtained as follows. 
 \begin{equation}
 \label{eq: DoF KUn0}
     \begin{aligned}
         \mathrm{DoF}= \frac{K_{M} \binom{P-1}{Q-1}\beta + K_{U} \left( 1-\gamma  \right) \binom{P}{\Brt} \beta^{\prime}}{\sum\limits_{r=1}^{P-Q+1} D\left( \delta_{r}\right) \binom{P-r}{Q-1}+\left\lceil \frac{K_{U} \left( 1-\gamma \right) \binom{P}{\Brt}\beta^{\prime}}{\min \left( K_{U},\alpha \right)} \right\rceil}.
     \end{aligned}
 \end{equation}
 Moreover, for $K_{U}=0$, we have $K_{M}=K$, and by applying \eqref{eq: JM A apx} and \eqref{eq: TM}  into \eqref{eq: DoF def}, the DoF takes the form as:
 \begin{equation}
 \label{eq: DoF KU0}
     \begin{aligned}
         \mathrm{DoF}=\frac{K \binom{P-1}{Q-1}\beta}{\sum\limits_{r=1}^{P-Q+1} D\left( \delta_{r}\right) \binom{P-r}{Q-1}}.
     \end{aligned}
 \end{equation}
\subsection{Proof of Theorem~\ref{thm: DoF} for \textit{Strategy B}}
\label{subsection: proof B}
 In order to obtain the DoF for the dynamic MISO network operating with \textit{Strategy B}, first, we count the number of times that users are served during the CC delivery step. In this regard, assume that user $k$ with $p \left[k \right]=p$ is present in the CC delivery step, and receives its subpackets with \textit{Strategy~B}. According to \eqref{eq: erm}, \eqref{eq: Krsmu} and \eqref{eq: Icr non-int}, when $c \in \left[ P-Q+1 \right]$, $l \in \left[ \binom{P-c-1}{Q-2} \right]$, $m \in \left[ \theta \right]$ and $s \in \left[ \nu_{2} \right]$ , user $k$ is served during the transmission quintuples $\left(p,c,l,m,s \right)$. Hence, for $r=p$, the number of transmission quintuples $\left( r,c,l,m,s \right)$ serving user $k$ is given by:
\begin{equation}
    \begin{aligned}
           \sum\limits_{c=1}^{P-Q+1} \binom{P-c-1}{Q-2} \theta\nu_{2},
    \end{aligned}
\end{equation}
which by using Lemma~\ref{lemma: comb} is simplified to:
\begin{equation}
\label{eq: ser1}
    \begin{aligned}
           \sum\limits_{c=1}^{P-Q+1} \binom{P-c-1}{Q-2} \theta\nu_{2}=\theta\nu_{2} \binom{P-1}{Q-1}.
    \end{aligned}
\end{equation}
Furthermore, as per \eqref{eq: erm}, \eqref{eq: Krsmu} and \eqref{eq: Icr non-int}, when $r \neq p$, $c \in \left[ P-Q+1 \right]$, $l \in \left[ \binom{P-c-2}{Q-3}  \right]$, $m \in \left[ \hat{\eta} \right]$ and $s \in \left[ \nu_{2} \right]$, the transmission quintuples $\left( r,c,l,m,s \right)$ serve user $k$. Therefore, the number of transmission quintuples $\left(r,c,l,m,s \right)$ serving user $k$ with $r\neq p$ is obtained as:
\begin{equation}
    \begin{aligned}
          \sum\limits_{r=1}^{P-1} \sum\limits_{c=1}^{P-Q+1} \binom{P-c-2}{Q-3} \hat{\eta} \nu_{2},
    \end{aligned}
\end{equation}
which by using Lemma~\ref{lemma: comb} takes the form as follows. 
\begin{equation}
\label{eq: ser2}
    \begin{aligned}
          \sum\limits_{r=1}^{P-1} \sum\limits_{c=1}^{P-Q+1} \binom{P-c-2}{Q-3} \hat{\eta} \nu_{2}=
         \hat{\eta} \nu_{2} \left( P-1 \right) \binom{P-2}{Q-2}.
    \end{aligned}
\end{equation}
As a result, by using \eqref{eq: ser1}, \eqref{eq: ser2} and according to that $\theta=\alpha-\hat{\eta} \left\lfloor \frac{\alpha}{\hat{\eta}} \right\rfloor$, the number of times that user $k$ is served with \textit{Strategy B} is equal to:
\begin{equation}
\label{eq: total service k}
    \begin{aligned}
          & \theta \nu_{2} \binom{P-1}{Q-1}+ \hat{\eta} \nu_{2} \left( P-1 \right) \binom{P-2}{Q-2}=\\
          &\binom{P-1}{Q-1} \nu_{2} \left( \hat{\eta} \Brt +\alpha \right).
    \end{aligned}
\end{equation}
Since $K_{M}$ users are present in the CC delivery step, the total number of times that all $K_{M}$ users are served with \textit{Strategy~B} is given by:
\begin{equation}
\label{eq: JM strategy B}
    \begin{aligned}
         J_{M}=K_{M}\binom{P-1}{Q-1} \nu_{2} \left( \hat{\eta} \Brt +\alpha \right).
    \end{aligned}
\end{equation}

In the proceeding, we count the total number of transmissions with \textit{Strategy B}. As stated in Section~\ref{subsection: strategy B}, if $I^{+} \left( \Bar{\delta}_{c}, \CE_{r}^{m} \right)=1$, the server delivers the subpackets to the users during the transmission quintuple $\left( r,c,l,m,s \right)$; otherwise, it does not transmit any signal. Now, by following a similar approach to \eqref{eq: TM},  since $r \in [P]$, $c \in \left[ P-Q+1 \right]$, $l \in \left[ \binom{P-c-1}{Q-2} \right]$, $ m \in \left[ \hat{\eta} \right]$ and $s \in \left[ \nu_{2} \right]$, the total number of transmissions in \textit{Strategy B} is given by:
\begin{equation}
  \label{eq: TM B}
      \begin{aligned}
       T_{M}= \sum_{r=1}^{P} \sum_{c=1}^{P-Q+1} \sum_{m=1}^{\hat{\eta}} \sum_{s=1}^{\nu_{2}} \binom{P-c-1}{Q-2}  I^{+} \left( \Bar{\delta}_{c}, \CE_{r}^{m} \right).
      \end{aligned}
  \end{equation}

At the next step, we show that user $k$ receives all its subpackets during the CC delivery step with \textit{Strategy B}. As observed from \eqref{eq: total service k}, each user $k$ is served for $\binom{P-1}{Q-1} \nu_{2} \left( \hat{\eta} \Brt +\alpha \right)$ times during the CC delivery step. As per \eqref{eq: erm}-\eqref{eq: Icr non-int}, in each transmission, user $k$ receives $\nu_{1}$  subpackets. Therefore, the total number of subpackets that user $k$ receives with \textit{Strategy B} is $\binom{P-1}{Q-1} \nu_{1} \nu_{2} \left( \hat{\eta} \Brt +\alpha \right)$, which by following the same way as in  \eqref{eq: sub int}, can be reformulated as follows. 
\begin{equation}
    \begin{aligned}
      & \binom{P-1}{Q-1} \binom{Q-2}{Q-\Brt-2} \binom{Q-1}{Q-\Brt-1} \left( \hat{\eta} \Brt +\alpha \right)=\\
      &\left( 1-\gamma\right) \binom{P}{\Brt}\binom{P-\Brt-1}{Q-\Brt-1} \binom{Q-2}{Q-\Brt-2} \left( \hat{\eta} \Brt +\alpha \right),   
    \end{aligned}
\end{equation}
 which shows that user $k$ receives all subpackets of its requested file during the CC delivery step with \textit{Strategy B}. 
 
Next, in order to obtain $J_{U}$ and $T_{U}$, we follow the same way as in \eqref{eq: JU A apx} and \eqref{eq: TU}. Accordingly, as stated in Section~\ref{subsection: strategy B}, each file is split into $\binom{P}{\Brt}\binom{P-\Brt-1}{Q-\Brt-1} \binom{Q-2}{Q-\Brt-2} \left( \hat{\eta} \Brt +\alpha \right)$ subpackets. Therefore, according to the fact that $K_{U}$ users are present in the UC delivery step and the cache ratio at each user is $\gamma$, the total number of times that users served during the UC delivery step is given by:
\begin{equation}
\label{eq: JU B}
    \begin{aligned}
         J_{U}&=K_{U} \left( 1-\gamma \right) \binom{P}{\Brt}\binom{P-\Brt-1}{Q-\Brt-1} \binom{Q-2}{Q-\Brt-2} \left( \hat{\eta} \Brt +\alpha \right)\\
         &=K_{U} \left( 1-\gamma  \right) \binom{P}{\Brt} \alpha^{\prime}.
    \end{aligned}
\end{equation}
Furthermore, as mentioned in Section~\ref{section: UC}, assuming $K_{U}\neq 0$, each transmission of the UC delivery step serves $\min \left(K_{U}, \alpha \right)$ users. Hence, the total number of transmissions in the UC delivery step is obtained as follows. 
\begin{equation}
\label{eq: TU B}
    \begin{aligned}
         T_{U}=\left\lceil \frac{K_{U} \left( 1-\gamma \right) \binom{P}{\Brt}\alpha^{\prime}}{\min \left( K_{U},\alpha \right)} \right\rceil.
    \end{aligned}
\end{equation}
As a result, for $K_{U}\neq 0$, by applying \eqref{eq: JM strategy B}, \eqref{eq: TM B}, \eqref{eq: JU B} and \eqref{eq: TU B} into \eqref{eq: DoF def}, the DoF is obtained as follows. 
\begin{equation}
    \begin{aligned}
          \mathrm{DoF}= \frac{K_{M} \binom{P-1}{Q-1}\left( \hat{\eta}\Brt+\alpha \right)\nu_{2} + K_{U} \left( 1-\gamma  \right) \binom{P}{\Brt} \alpha^{\prime}}{N_{M}+N_{U}}.
    \end{aligned}
\end{equation}
In addition, when $K_{U}=0$ and $K_{M}=K$, by substituting \eqref{eq: JM strategy B} and \eqref{eq: TM B} into \eqref{eq: DoF def}, the DoF is simplified to:
\begin{equation}
\label{eq: DoF Ku0 B}
    \mathrm{DoF}=  \frac{K \binom{P-1}{Q-1}\left( \hat{\eta}\Brt+\alpha \right) \nu_{2}}{N_{M}}.
\end{equation}

\section{Proof of Remark~\ref{rem: optimal}}
\label{apx: proof DoF alphaleta uniform}
When $K$ users are uniformly assigned to $P$ caching profiles, i.e., $K=P\hat{\eta}$, and all users are served via CC delivery step, i.e., $K_{M}=K$ and $K_{U}=0$, we prove Remark~\ref{rem: optimal} for three regimes: \textit{1)} $\alpha \leq \hat{\eta}$, \textit{2)} $\alpha> \hat{\eta}$ with integer $\frac{\alpha}{\hat{\eta}}$ and \textit{3)} $\alpha> \hat{\eta}$ with non-integer $\frac{\alpha}{\hat{\eta}}$.

\textit{1)}  $\alpha\leq \hat{\eta}$: In this regime, the system operates with \textit{Strategy A} during the CC delivery step, where  $\beta=\alpha$ and $Q=\Brt+1$. By considering $\hat{\eta}=\eta_{r}$ for $\forall r \in [P]$, we know that $D\left( \delta_{r} \right)=\hat{\eta}$. Next. we reformulate \eqref{eq: DoF thm} as follows. 
    \begin{equation}
\label{eq: DoF alpha l eta simp}
    \begin{aligned}
    \mathrm{DoF}=\frac{K \binom{P-1}{Q-1}\alpha}{ \hat{\eta} \sum\limits_{r=1}^{P-Q+1} \binom{P-r}{Q-1}}.
    \end{aligned}
\end{equation}
   Now, using Lemma~\ref{lemma: comb} and setting $Q=\Brt +1$, the DoF in \eqref{eq: DoF alpha l eta simp} is simplified to:
    \begin{equation}
\label{eq: DoF alpha l eta final}
    \begin{aligned}
    \mathrm{DoF}=\frac{K \binom{P-1}{Q-1}\alpha}{ \hat{\eta} \binom{P}{Q}}= \frac{K Q \alpha}{P\hat{\eta}}=\alpha \left( P\gamma +1 \right).
    \end{aligned}
\end{equation}

\textit{2)}  $\alpha > \hat{\eta}$ and integer $\frac{\alpha}{\hat{\eta}}$: In this regime, we know that the system uses \textit{Strategy A} to deliver data during the CC delivery step, where $\beta=\hat{\eta}$ and $\Brt +1 \leq Q \leq \Brt + \frac{\alpha}{\hat{\eta}}$. Here, in order to maximize the achievable DoF, we set $Q=\Brt + \frac{\alpha}{\hat{\eta}}$. Then, similarly to the previous case, we consider $\hat{\eta}=\eta_{r}$ and $D \left( \delta_{r} \right)=\hat{\eta}$ for all $r \in [P]$. Accordingly, the DoF in \eqref{eq: DoF thm} is given by:
 \begin{equation}
\label{eq: DoF alpha g eta uni}
    \begin{aligned}
    \mathrm{DoF}=\frac{K \binom{P-1}{Q-1}\hat{\eta}}{ \hat{\eta} \sum\limits_{r=1}^{P-Q+1} \binom{P-r}{Q-1}}.
    \end{aligned}
\end{equation}
Finally, by using Lemma~\ref{lemma: comb} and \eqref{eq: DoF alpha g eta uni}, the maximum achievable DoF corresponding to this regime is as follows. 
\begin{equation}
    \begin{aligned}
    \mathrm{DoF}=\nicefrac{K \binom{P-1}{Q-1}}{\binom{P}{Q}}=
    \nicefrac{KQ}{P}=
    \hat{\eta}\Brt+\alpha=
    K\gamma+\alpha.
    \end{aligned}
\end{equation}

\textit{3)} $\alpha> \hat{\eta}$ with non-integer $\frac{\alpha}{\hat{\eta}}$: In this regime, in order to maximize the achievable DoF, the server can transmit data via \textit{Strategy B} during the CC delivery step, such that $\hat{\eta}=\eta_{r}$ for all $r \in [P]$, $\beta=\hat{\eta}$ and $Q=\Brt+\left\lceil \frac{\alpha}{\hat{\eta}} \right\rceil$. Accordingly, we have $I^{+} \left( \Bar{\delta_{c}}, \CE_{r}^{m} \right)=1$ for all $r \in [P]$, $c \in \left[P-Q+1 \right]$, $l \in \left[ \binom{P-c-1}{Q-2} \right]$, $m \in \left[ \hat{\eta} \right]$ and $s \in \left[ \nu_{2} \right]$. Hence, by using Lemma~\ref{lemma: comb} and \eqref{eq: NM def B}, $N_{M}$ is simplified to:
\begin{equation}
\label{eq: NM uniform B}
    N_{M}=P\hat{\eta} \binom{P-1}{Q-1}\nu_{2}.
\end{equation}
Finally, by substituting \eqref{eq: NM uniform B} into \eqref{eq: DoF Ku0 B} and using $\Brt=P\gamma$, the DoF of this regime for the uniform user-to-profile association takes the form of:
\begin{equation}
    \begin{aligned}
    \label{eq: DoF final uniform B}
         \mathrm{DoF}=\frac{K\binom{P-1}{Q-1}\left( \hat{\eta} \Brt +\alpha \right) \nu_{2}}{P\hat{\eta} \binom{P-1}{Q-1}\nu_{2}}= \hat{\eta} \Brt+\alpha=K\gamma+\alpha.
   \end{aligned}
\end{equation}


\section{Proof of Remark~\ref{rem: DoF max alphaleta}}
\label{apx: Proof DoF max alphaleta}
   For $Q=\Brt+1$ and $\eta_{1} \leq \alpha$, the system operates with $\textit{Strategy~A}$ during the CC delivery step, and $\beta^{\prime}=\beta$, where $\beta=\alpha$, if $\alpha \leq \hat{\eta}$; otherwise, $\beta=\hat{\eta}$. Assuming $\hat{\eta} < \eta_{1}$ (i.e. $K_{U}>0$) and using \eqref{eq: TU}, clearly, it is sure that $ \binom{P}{Q} + \frac{1}{\beta} T_{U} >\binom{P}{Q} $. So, by using Lemma~\ref{lemma: comb}, we have:
    \begin{equation}
    \label{eq: ineq tx alpha g eta def2}
        \begin{aligned}
      &\frac{1}{\beta} \sum\limits_{r=1}^{P-Q+1} \beta \binom{P-r}{Q-1} +\frac{1}{\beta}T_{U} >  
      \frac{1}{\eta_{1}} \sum\limits_{r=1}^{P-Q+1} \eta_{1} \binom{P-r}{Q-1}.
        \end{aligned}
    \end{equation} 
    Now, according to the fact that for any arbitrary parameters $w$, $x$, $y$ and $z$, if $w+x \geq y$, then $\max \left( w,z \right)+x \geq y$, we  rearrange \eqref{eq: ineq tx alpha g eta def2} as follows. 
    \begin{equation}
    \small
    \label{eq: ineq tx alpha g eta def}
        \begin{aligned}
      &\frac{1}{\beta} \sum\limits_{r=1}^{P-Q+1} \max \left( \beta, \delta_{r} \right) \binom{P-r}{Q-1} +\frac{T_{U}}{\beta}  >
      \frac{1}{\eta_{1}} \sum\limits_{r=1}^{P-Q+1} \eta_{1} \binom{P-r}{Q-1}.
        \end{aligned}
    \end{equation} 
    We know that $D \left( \delta_{r} \right)=D \left( \eta_{r} \right)=0$ if $\eta_{r}=0$; otherwise, $D \left( \delta_{r} \right)=\max \left( \beta , \delta_{r} \right)$ and $D \left( \eta_{r} \right)=\max \left( \eta_{r}, \eta_{1} \right)=\eta_{1}$. Hence, we reformulate \eqref{eq: ineq tx alpha g eta def} as follows. 
    \begin{equation}
    \label{eq: ineq tx alpha g eta simp}
        \begin{aligned}
      &\frac{ K \binom{P-1}{Q-1} \beta}{\sum\limits_{r=1}^{P-Q+1} D \left( \delta_{r} \right) \binom{P-r}{Q-1} +T_{U} }  < 
       \frac{K \binom{P-1}{Q-1} \eta_{1}}{\sum\limits_{r=1}^{P-Q+1} D \left( \eta_{r} \right) \binom{P-r}{Q-1}}.
        \end{aligned}
    \end{equation}
    Furthermore, using $\beta^{\prime}=\beta$, $Q=\Brt+1$ and \eqref{eq: JM plus JU}, we have:
    \begin{equation}
    \label{eq: subpacket rem}
        \begin{aligned}
        J_{M}+J_{U}=K\binom{P-1}{Q-1}\beta.
        \end{aligned}
    \end{equation}
    According to \eqref{eq: subpacket rem} and \eqref{eq: DoF thm},  the left side of \eqref{eq: ineq tx alpha g eta simp} demonstrates the DoF of a setup with $Q=\Brt+1$ and $\hat{\eta}<\eta_{1}$ (i.e., $K_{U} \neq 0$), while the right side corresponds to the DoF with $Q=\Brt+1$ and $\hat{\eta}=\eta_{1}$ (i.e., $K_{U} = 0$). As a result, for $Q=\Brt+1$ and $\eta_{1} \leq \alpha$, we should set $\hat{\eta}=\eta_{1}$ to maximize the achievable DoF.



\end{document}

%% file: commands.tex

\newtheorem{lemma}{Lemma}
\newtheorem{corollary}{Corollary}

\newcommand{\diag}{{\mbox{diag}}}
\newcommand{\herm}{^{\mbox{\scriptsize H}}}
\newcommand{\sherm}{^{\mbox{\scriptsize H}}}
\newcommand{\tran}{^{\mbox{\scriptsize T}}}
\newcommand{\stran}{^{\mbox{\tiny T}}}

\newcommand{\vbar}{\raisebox{.17ex}{\rule{.04em}{1.35ex}}}
\newcommand{\vbarind}{\raisebox{.01ex}{\rule{.04em}{1.1ex}}}
\newcommand{\R}{\ifmmode{\rm I}\hspace{-.2em}{\rm R} \else ${\rm I}\hspace{-.2em}{\rm R}$ \fi}
\newcommand{\T}{\ifmmode{\rm I}\hspace{-.2em}{\rm T} \else ${\rm I}\hspace{-.2em}{\rm T}$ \fi}
\newcommand{\N}{\ifmmode{\rm I}\hspace{-.2em}{\rm N} \else \mbox{${\rm I}\hspace{-.2em}{\rm N}$} \fi}
\newcommand{\B}{\ifmmode{\rm I}\hspace{-.2em}{\rm B} \else \mbox{${\rm I}\hspace{-.2em}{\rm B}$} \fi}
\newcommand{\Hil}{\ifmmode{\rm I}\hspace{-.2em}{\rm H} \else \mbox{${\rm I}\hspace{-.2em}{\rm H}$} \fi}
\newcommand{\C}{\ifmmode\hspace{.2em}\vbar\hspace{-.31em}{\rm C} \else \mbox{$\hspace{.2em}\vbar\hspace{-.31em}{\rm C}$} \fi}
\newcommand{\Cind}{\ifmmode\hspace{.2em}\vbarind\hspace{-.25em}{\rm C} \else \mbox{$\hspace{.2em}\vbarind\hspace{-.25em}{\rm C}$} \fi}
\newcommand{\Q}{\ifmmode\hspace{.2em}\vbar\hspace{-.31em}{\rm Q} \else \mbox{$\hspace{.2em}\vbar\hspace{-.31em}{\rm Q}$} \fi}
\newcommand{\Z}{\ifmmode{\rm Z}\hspace{-.28em}{\rm Z} \else ${\rm Z}\hspace{-.28em}{\rm Z}$ \fi}

\newcommand{\sgn}{\mbox{sgn}}
\newcommand{\var}{\mbox{var}}
\renewcommand{\Re}{\mbox{Re}}
\renewcommand{\Im}{\mbox{Im}}

\renewcommand{\vec}[1]{\bf{#1}}     
\newcommand{\vecsc}[1]{\mbox{\bf \scriptsize #1}}
\newcommand{\itvec}[1]{\mbox{\boldmath{$#1$}}}
\newcommand{\itvecsc}[1]{\mbox{\boldmath{$\scriptstyle #1$}}}
\newcommand{\gvec}[1]{\mbox{\boldmath{$#1$}}}

\newcommand{\balpha}{\mbox{\boldmath{$\alpha$}}}
\newcommand{\bbeta}{\mbox{\boldmath{$\beta$}}}
\newcommand{\bgamma}{\mbox{\boldmath{$\gamma$}}}
\newcommand{\bdelta}{\mbox{\boldmath{$\delta$}}}
\newcommand{\bepsilon}{\mbox{\boldmath{$\epsilon$}}}
\newcommand{\bvarepsilon}{\mbox{\boldmath{$\varepsilon$}}}
\newcommand{\bzeta}{\mbox{\boldmath{$\zeta$}}}
\newcommand{\boldeta}{\mbox{\boldmath{$\eta$}}}
\newcommand{\btheta}{\mbox{\boldmath{$\theta$}}}
\newcommand{\bvartheta}{\mbox{\boldmath{$\vartheta$}}}
\newcommand{\biota}{\mbox{\boldmath{$\iota$}}}
\newcommand{\blambda}{\mbox{\boldmath{$\lambda$}}}
\newcommand{\bmu}{\mbox{\boldmath{$\mu$}}}
\newcommand{\bnu}{\mbox{\boldmath{$\nu$}}}
\newcommand{\bxi}{\mbox{\boldmath{$\xi$}}}
\newcommand{\bpi}{\mbox{\boldmath{$\pi$}}}
\newcommand{\bvarpi}{\mbox{\boldmath{$\varpi$}}}
\newcommand{\brho}{\mbox{\boldmath{$\rho$}}}
\newcommand{\bvarrho}{\mbox{\boldmath{$\varrho$}}}
\newcommand{\bsigma}{\mbox{\boldmath{$\sigma$}}}
\newcommand{\bvarsigma}{\mbox{\boldmath{$\varsigma$}}}
\newcommand{\btau}{\mbox{\boldmath{$\tau$}}}
\newcommand{\bupsilon}{\mbox{\boldmath{$\upsilon$}}}
\newcommand{\bphi}{\mbox{\boldmath{$\phi$}}}
\newcommand{\bvarphi}{\mbox{\boldmath{$\varphi$}}}
\newcommand{\bchi}{\mbox{\boldmath{$\chi$}}}
\newcommand{\bpsi}{\mbox{\boldmath{$\psi$}}}
\newcommand{\bomega}{\mbox{\boldmath{$\omega$}}}

\newcommand{\bolda}{\mbox{\boldmath{$a$}}}
\newcommand{\bb}{\mbox{\boldmath{$b$}}}
\newcommand{\bc}{\mbox{\boldmath{$c$}}}
\newcommand{\bd}{\mbox{\boldmath{$d$}}}
\newcommand{\bolde}{\mbox{\boldmath{$e$}}}
\newcommand{\boldf}{\mbox{\boldmath{$f$}}}
\newcommand{\bg}{\mbox{\boldmath{$g$}}}
\newcommand{\bh}{\mbox{\boldmath{$h$}}}
\newcommand{\bp}{\mbox{\boldmath{$p$}}}
\newcommand{\bq}{\mbox{\boldmath{$q$}}}
\newcommand{\br}{\mbox{\boldmath{$r$}}}
\newcommand{\bs}{\mbox{\boldmath{$s$}}}
\newcommand{\bt}{\mbox{\boldmath{$t$}}}
\newcommand{\bu}{\mbox{\boldmath{$u$}}}
\newcommand{\bv}{\mbox{\boldmath{$v$}}}
\newcommand{\bw}{\mbox{\boldmath{$w$}}}
\newcommand{\bx}{\mbox{\boldmath{$x$}}}
\newcommand{\by}{\mbox{\boldmath{$y$}}}
\newcommand{\bz}{\mbox{\boldmath{$z$}}}

%% file: MyCommands.tex
\newtheorem{thm}{Theorem}
\newtheorem{lem}{Lemma}
\newtheorem{prop}[thm]{Proposition}
\newtheorem{cor}{Corollary}
\newtheorem{conj}{Conjecture}[section]
\newtheorem{exmp}{Example}
\newtheorem{defn}{Definition}

\newtheorem{rem}{Remark}

\newcommand{\CA}[0]{{\mathcal{A}}}
\newcommand{\CB}[0]{{\mathcal{B}}}
\newcommand{\CC}[0]{{\mathcal{C}}}
\newcommand{\CD}[0]{{\mathcal{D}}}
\newcommand{\CE}[0]{{\mathcal{E}}}
\newcommand{\CF}[0]{{\mathcal{F}}}
\newcommand{\CG}[0]{{\mathcal{G}}}
\newcommand{\CH}[0]{{\mathcal{H}}}
\newcommand{\CI}[0]{{\mathcal{I}}}
\newcommand{\CJ}[0]{{\mathcal{J}}}
\newcommand{\CK}[0]{{\mathcal{K}}}
\newcommand{\CL}[0]{{\mathcal{L}}}
\newcommand{\CM}[0]{{\mathcal{M}}}
\newcommand{\CN}[0]{{\mathcal{N}}}
\newcommand{\CO}[0]{{\mathcal{O}}}
\newcommand{\CP}[0]{{\mathcal{P}}}
\newcommand{\CQ}[0]{{\mathcal{Q}}}
\newcommand{\CR}[0]{{\mathcal{R}}}
\newcommand{\CS}[0]{{\mathcal{S}}}
\newcommand{\CT}[0]{{\mathcal{T}}}
\newcommand{\CU}[0]{{\mathcal{U}}}
\newcommand{\CV}[0]{{\mathcal{V}}}
\newcommand{\CW}[0]{{\mathcal{W}}}
\newcommand{\CX}[0]{{\mathcal{X}}}
\newcommand{\CY}[0]{{\mathcal{Y}}}
\newcommand{\CZ}[0]{{\mathcal{Z}}}

\newcommand{\Ba}[0]{{\mathbf{a}}}
\newcommand{\Bb}[0]{{\mathbf{b}}}
\newcommand{\Bc}[0]{{\mathbf{c}}}
\newcommand{\Bd}[0]{{\mathbf{d}}}
\newcommand{\Be}[0]{{\mathbf{e}}}
\newcommand{\Bf}[0]{{\mathbf{f}}}
\newcommand{\Bg}[0]{{\mathbf{g}}}
\newcommand{\Bh}[0]{{\mathbf{h}}}
\newcommand{\Bi}[0]{{\mathbf{i}}}
\newcommand{\Bj}[0]{{\mathbf{j}}}
\newcommand{\Bk}[0]{{\mathbf{k}}}
\newcommand{\Bl}[0]{{\mathbf{l}}}
\newcommand{\Bm}[0]{{\mathbf{m}}}
\newcommand{\Bn}[0]{{\mathbf{n}}}
\newcommand{\Bo}[0]{{\mathbf{o}}}
\newcommand{\Bp}[0]{{\mathbf{p}}}
\newcommand{\Bq}[0]{{\mathbf{q}}}
\newcommand{\Br}[0]{{\mathbf{r}}}
\newcommand{\Bs}[0]{{\mathbf{s}}}
\newcommand{\Bt}[0]{{\mathbf{t}}}
\newcommand{\Bu}[0]{{\mathbf{u}}}
\newcommand{\Bv}[0]{{\mathbf{v}}}
\newcommand{\Bw}[0]{{\mathbf{w}}}
\newcommand{\Bx}[0]{{\mathbf{x}}}
\newcommand{\By}[0]{{\mathbf{y}}}
\newcommand{\Bz}[0]{{\mathbf{z}}}

\newcommand{\BA}[0]{{\mathbf{A}}}
\newcommand{\BB}[0]{{\mathbf{B}}}
\newcommand{\BC}[0]{{\mathbf{C}}}
\newcommand{\BD}[0]{{\mathbf{D}}}
\newcommand{\BE}[0]{{\mathbf{E}}}
\newcommand{\BF}[0]{{\mathbf{F}}}
\newcommand{\BG}[0]{{\mathbf{G}}}
\newcommand{\BH}[0]{{\mathbf{H}}}
\newcommand{\BI}[0]{{\mathbf{I}}}
\newcommand{\BJ}[0]{{\mathbf{J}}}
\newcommand{\BK}[0]{{\mathbf{K}}}
\newcommand{\BL}[0]{{\mathbf{L}}}
\newcommand{\BM}[0]{{\mathbf{M}}}
\newcommand{\BN}[0]{{\mathbf{N}}}
\newcommand{\BO}[0]{{\mathbf{O}}}
\newcommand{\BP}[0]{{\mathbf{P}}}
\newcommand{\BQ}[0]{{\mathbf{Q}}}
\newcommand{\BR}[0]{{\mathbf{R}}}
\newcommand{\BS}[0]{{\mathbf{S}}}
\newcommand{\BT}[0]{{\mathbf{T}}}
\newcommand{\BU}[0]{{\mathbf{U}}}
\newcommand{\BV}[0]{{\mathbf{V}}}
\newcommand{\BW}[0]{{\mathbf{W}}}
\newcommand{\BX}[0]{{\mathbf{X}}}
\newcommand{\BY}[0]{{\mathbf{Y}}}
\newcommand{\BZ}[0]{{\mathbf{Z}}}

\newcommand{\Bra}[0]{{\Bar{a}}}
\newcommand{\Brb}[0]{{\Bar{b}}}
\newcommand{\Brc}[0]{{\Bar{c}}}
\newcommand{\Brd}[0]{{\Bar{d}}}
\newcommand{\Bre}[0]{{\Bar{e}}}
\newcommand{\Brf}[0]{{\Bar{f}}}
\newcommand{\Brg}[0]{{\Bar{g}}}
\newcommand{\Brh}[0]{{\Bar{h}}}
\newcommand{\Bri}[0]{{\Bar{i}}}
\newcommand{\Brj}[0]{{\Bar{j}}}
\newcommand{\Brk}[0]{{\Bar{k}}}
\newcommand{\Brl}[0]{{\Bar{l}}}
\newcommand{\Brm}[0]{{\Bar{m}}}
\newcommand{\Brn}[0]{{\Bar{n}}}
\newcommand{\Bro}[0]{{\Bar{o}}}
\newcommand{\Brp}[0]{{\Bar{p}}}
\newcommand{\Brq}[0]{{\Bar{q}}}
\newcommand{\Brr}[0]{{\Bar{r}}}
\newcommand{\Brs}[0]{{\Bar{s}}}
\newcommand{\Brt}[0]{{\Bar{t}}}
\newcommand{\Bru}[0]{{\Bar{u}}}
\newcommand{\Brv}[0]{{\Bar{v}}}
\newcommand{\Brw}[0]{{\Bar{w}}}
\newcommand{\Brx}[0]{{\Bar{x}}}
\newcommand{\Bry}[0]{{\Bar{y}}}
\newcommand{\Brz}[0]{{\Bar{z}}}

\newcommand{\BrA}[0]{{\Bar{A}}}
\newcommand{\BrB}[0]{{\Bar{B}}}
\newcommand{\BrC}[0]{{\Bar{C}}}
\newcommand{\BrD}[0]{{\Bar{D}}}
\newcommand{\BrE}[0]{{\Bar{E}}}
\newcommand{\BrF}[0]{{\Bar{F}}}
\newcommand{\BrG}[0]{{\Bar{G}}}
\newcommand{\BrH}[0]{{\Bar{H}}}
\newcommand{\BrI}[0]{{\Bar{I}}}
\newcommand{\BrJ}[0]{{\Bar{J}}}
\newcommand{\BrK}[0]{{\Bar{K}}}
\newcommand{\BrL}[0]{{\Bar{L}}}
\newcommand{\BrM}[0]{{\Bar{M}}}
\newcommand{\BrN}[0]{{\Bar{N}}}
\newcommand{\BrO}[0]{{\Bar{O}}}
\newcommand{\BrP}[0]{{\Bar{P}}}
\newcommand{\BrQ}[0]{{\Bar{Q}}}
\newcommand{\BrR}[0]{{\Bar{R}}}
\newcommand{\BrS}[0]{{\Bar{S}}}
\newcommand{\BrT}[0]{{\Bar{T}}}
\newcommand{\BrU}[0]{{\Bar{U}}}
\newcommand{\BrV}[0]{{\Bar{V}}}
\newcommand{\BrW}[0]{{\Bar{W}}}
\newcommand{\BrX}[0]{{\Bar{X}}}
\newcommand{\BrY}[0]{{\Bar{Y}}}
\newcommand{\BrZ}[0]{{\Bar{Z}}}

\newcommand{\Sfa}[0]{{\mathsf{a}}}
\newcommand{\Sfb}[0]{{\mathsf{b}}}
\newcommand{\Sfc}[0]{{\mathsf{c}}}
\newcommand{\Sfd}[0]{{\mathsf{d}}}
\newcommand{\Sfe}[0]{{\mathsf{e}}}
\newcommand{\Sff}[0]{{\mathsf{f}}}
\newcommand{\Sfg}[0]{{\mathsf{g}}}
\newcommand{\Sfh}[0]{{\mathsf{h}}}
\newcommand{\Sfi}[0]{{\mathsf{i}}}
\newcommand{\Sfj}[0]{{\mathsf{j}}}
\newcommand{\Sfk}[0]{{\mathsf{k}}}
\newcommand{\Sfl}[0]{{\mathsf{l}}}
\newcommand{\Sfm}[0]{{\mathsf{m}}}
\newcommand{\Sfn}[0]{{\mathsf{n}}}
\newcommand{\Sfo}[0]{{\mathsf{o}}}
\newcommand{\Sfp}[0]{{\mathsf{p}}}
\newcommand{\Sfq}[0]{{\mathsf{q}}}
\newcommand{\Sfr}[0]{{\mathsf{r}}}
\newcommand{\Sfs}[0]{{\mathsf{s}}}
\newcommand{\Sft}[0]{{\mathsf{t}}}
\newcommand{\Sfu}[0]{{\mathsf{u}}}
\newcommand{\Sfv}[0]{{\mathsf{v}}}
\newcommand{\Sfw}[0]{{\mathsf{w}}}
\newcommand{\Sfx}[0]{{\mathsf{x}}}
\newcommand{\Sfy}[0]{{\mathsf{y}}}
\newcommand{\Sfz}[0]{{\mathsf{z}}}

\newcommand{\SfA}[0]{{\mathsf{A}}}
\newcommand{\SfB}[0]{{\mathsf{B}}}
\newcommand{\SfC}[0]{{\mathsf{C}}}
\newcommand{\SfD}[0]{{\mathsf{D}}}
\newcommand{\SfE}[0]{{\mathsf{E}}}
\newcommand{\SfF}[0]{{\mathsf{F}}}
\newcommand{\SfG}[0]{{\mathsf{G}}}
\newcommand{\SfH}[0]{{\mathsf{H}}}
\newcommand{\SfI}[0]{{\mathsf{I}}}
\newcommand{\SfJ}[0]{{\mathsf{J}}}
\newcommand{\SfK}[0]{{\mathsf{K}}}
\newcommand{\SfL}[0]{{\mathsf{L}}}
\newcommand{\SfM}[0]{{\mathsf{M}}}
\newcommand{\SfN}[0]{{\mathsf{N}}}
\newcommand{\SfO}[0]{{\mathsf{O}}}
\newcommand{\SfP}[0]{{\mathsf{P}}}
\newcommand{\SfQ}[0]{{\mathsf{Q}}}
\newcommand{\SfR}[0]{{\mathsf{R}}}
\newcommand{\SfS}[0]{{\mathsf{S}}}
\newcommand{\SfT}[0]{{\mathsf{T}}}
\newcommand{\SfU}[0]{{\mathsf{U}}}
\newcommand{\SfV}[0]{{\mathsf{V}}}
\newcommand{\SfW}[0]{{\mathsf{W}}}
\newcommand{\SfX}[0]{{\mathsf{X}}}
\newcommand{\SfY}[0]{{\mathsf{Y}}}
\newcommand{\SfZ}[0]{{\mathsf{Z}}}


\renewcommand{\Re}{\mbox{Re}}
\renewcommand{\Im}{\mbox{Im}}

\renewcommand{\vec}[1]{\bf{#1}}     

\newcommand{\FillGray}[3]{\filldraw[gray!50](#3-1+0.1,#1-#2+0.1) rectangle (#3-0.1,#1-#2+1-0.1)}
\newcommand{\FillBlack}[3]{\filldraw[black!70](#3-1+0.1,#1-#2+0.1) rectangle (#3-0.1,#1-#2+1-0.1)}
\newcommand{\FillHatch}[3]{\fill[pattern=crosshatch, pattern color=black!65](#3-1,#1-#2)rectangle(#3,#1-#2+1)}
\ExplSyntaxOn

\NewDocumentCommand \vect { s o m }
 {
  \IfBooleanTF {#1}
   { \vectaux*{#3} }
   { \IfValueTF {#2} { \vectaux[#2]{#3} } { \vectaux{#3} } }
 }
\DeclarePairedDelimiterX \vectaux [1] {\lbrack} {\rbrack}
 { \, \dbacc_vect:n { #1 } \, }
\cs_new_protected:Npn \dbacc_vect:n #1
 {
  \seq_set_split:Nnn \l_tmpa_seq { , } { #1 }
  \seq_use:Nn \l_tmpa_seq { \enspace }
 }
\ExplSyntaxOff